\documentclass[useAMS,usenatbib]{mn2e}

\usepackage{mathptmx}
\usepackage{graphicx}
\usepackage{subfigure}
\usepackage{lpic}
\usepackage{multirow}
\newcommand\apjl{{ApJ}}%
\newcommand\apjs{{ApJS}}%
\newcommand\ao{{Appl.~Opt.}}%
\newcommand\aap{{A\&A}}%
\title[Strong lensing in X-ray vs. optical clusters]{The lensing
  efficiencies of MACS X-ray selected versus RCS optically selected galaxy clusters}

\author[Horesh et al.]{A.~Horesh,$^{1}$\thanks{E-mail:
    assafh@wise.tau.ac.il} D.~Maoz,$^{1}$ H.~Ebeling,$^{2}$ G.~Seidel,$^{3}$ and M.~Bartelmann$^{3}$\\
  $^{1}$School of Physics and Astronomy, Tel Aviv University, Tel
  Aviv 69978, Israel\\
  $^{2}$Institute for Astronomy, University of Hawaii, 2680 Woodlawn Drive, Honolulu, HI 96822, USA\\
  $^{3}$Zentrum f\"ur Astronomie der Universit\"at Heidelberg, Institut f\"ur Theoretische Astrophysik, Albert-\"Ueberle-Str. 2, 69120 Heidelberg, Germany\\}

\def\LaTeX{L\kern-.36em\raise.3ex\hbox{a}\kern-.15em
    T\kern-.1667em\lower.7ex\hbox{E}\kern-.125emX}
 
\begin{document}

\maketitle
\label{firstpage}
\begin{abstract}
  
  The statistics of strongly lensed arcs in samples of galaxy clusters
  provide information on cluster structure that is complementary to
  that from individual clusters. However, samples of clusters that
  have been analyzed to date have been either small, heterogeneous, or
  observed with limited angular resolution. We measure the lensed-arc
  statistics of 97 clusters imaged at high angular resolution with the
  Hubble Space Telescope, identifying lensed arcs using two automated
  arc detection algorithms. The sample includes similar numbers of
  X-ray selected (MACS) and optically selected (RCS) clusters, 
and spans cluster
  redshifts in the range $0.2 < z < 1$. We compile a catalogue of $42$
  arcs in the X-ray selected subsample and $7$ arcs in the optical
  subsample. All but five of these arcs are reported here for the
  first time. At $0.3 \leq z \leq 0.7$, the X-ray selected clusters
  have a significantly higher mean frequency of arcs, $1.2\pm 0.2$ per
  cluster, versus $0.2\pm 0.1$ in the optical sample. 
  The strikingly different lensing efficiencies indicate that X-ray
  clusters trace much larger mass concentrations, despite the
  similar optical luminosities of the X-ray and optical clusters. 
The mass difference is supported  also by 
  the lower space density of the X-ray clusters, 
and by the small Einstein
  radii of the few arcs in the optical sample. 
  Higher-order effects, such as differences in concentration or
  substructure, may also contribute.

\end{abstract}

\begin{keywords}
gravitational lensing -- galaxies: clusters: general
\end{keywords} 

\section{Introduction} 

Galaxy clusters are natural laboratories for studying a variety of
astrophysical processes and for testing cosmological models. In
particular, the masses and mass profiles of clusters have proved to be
useful for constraining cosmological parameters (e.g. Bridle et al.
1999; Reiprich \& B{\" o}hringer 2002; Voit 2005; Allen et al. 2008;
Vikhlinin et al. 2009). Gravitational lensing is frequently used to
map the evolution of cluster mass profiles, ellipticities, and
substructure. One approach is to perform detailed modeling of
individual clusters using strong and weak lensing (e.g., Abdelsalam et
al. 1998; Broadhurst et al.  2005; Leonard et al.  2007; Limousin et
al. 2007; Richard et al. 2007). However, since this kind of approach
requires deep data for individual clusters that exhibit numerous
lensed images, the results may not be representative of the vast
majority of clusters. A complementary approach is to measure the
statistics of lensed arcs in large samples of clusters.  Lensing
statistics thus provide another means to study clusters as a
population.

For the past decade there has been debate concerning theoretical
lensing statistics predictions and their confrontation with
observations. Bartelmann et al. (1998; B98) performed lensing
simulations using artificial sources at redshift $z=1$ by ray tracing
through the five most massive clusters formed in a cosmological N-body
dark matter simulation (Kauffmann et al. 1999). The observed number of
giant arcs, with length-to-width ratio $l/w \geq 10$ and $R<21.5$ mag, present
over the whole sky was estimated by extrapolating from observations of
a subsample of X-ray selected clusters from the Einstein Extended
Medium Sensitivity Survey (EMSS), and compared to the theoretical
calculation. B98 found that the estimated number of observed arcs is
larger by almost an order of magnitude than the number predicted by
the now-standard $\Lambda {\rm CDM}$ model. Later estimates of lensed
arcs statistics in clusters from both the Las Campanas Distant Cluster
Survey (Zaritsky \& Gonzales 2003; arcs with $l/w \geq 10$ and
$R<21.5$ mag) and the Red-Sequence Cluster Survey (RCS; Gladders et al.
2003) confirmed the estimates of the observed number of arcs derived
by B98. Most recently, Hennawi et al. (2008) analyzed a sample of 240
clusters, optically selected from the Sloan Digital Sky Survey (SDSS),
and found that $10\% - 20\%$ of them are strong lenses, similar to the
findings of Gladders et al. (2003). The largest catalogue of arcs to
date was compiled by Sand et al. (2005) who found $104$ arcs in $128$
clusters.  However, their systematic search for arcs was performed on
a largely heterogeneous cluster sample.

The apparent overproduction of arcs by real clusters has stimulated
further theoretical studies of arc statistics. Meneghetti et al.
(2000) studied numerically the effect of the masses of the individual
cluster galaxies on a cluster's lensing cross section, and found it to
be negligible, as also found in a study by Flores, Maller, \& Primack
(2000). However, the increase in lensing cross section due to the
central cluster cD galaxy may be as high as $\sim 50\%$ (Meneghetti,
Bartelmann, \& Moscardini 2003) and the increase in cross section due
to the intra-cluster gas could perhaps be by a factor of a few
(Puchwein et al.  2005, Rozo et al.  2008). Oguri, Lee, \& Suto (2003)
argued that halo triaxiality could also play an important role in
increasing cluster lensing cross sections.  Torri et al. (2004) raised
the possibility that X-ray selection of clusters may favor merging
systems, which may be more efficient lenses. Wambsganss, Bode, \&
Ostriker (2004) pointed out that since lensing cross section is a
steep function of source redshift, the conflict between theory and
observations could be the result of the assumed source redshifts in
the simulations. Similarly, Dalal, Holder, \& Hennawi (2004) performed
a lensing simulation using artificial background sources at different
redshifts and a large sample of simulated clusters.  They found that
their prediction for the number of lensed arcs was consistent with an
observed number that they derived from a sample of X-ray selected EMSS
clusters. The difference between this result and that of B98 was
explained by the combination of three main effects: the inclusion of
sources at different redshifts; the use of a higher source density in
the Dalal et al.  simulation; and an observed cluster number density
lower than the one used by B98 for estimating the all-sky number of
arcs.

A more observationally oriented approach to lensing statistics
simulations was introduced by Horesh et al. (2005; H05) in order to
test specifically the lensing efficiency of individual clusters,
independent of the separate question of the number density of
clusters. H05 repeated the B98 simulations using the same simulated
clusters, but using background sources from the Hubble Deep Field
(HDF), each at a redshift based on its actual photometric redshift.
Observational effects including background, photon noise, and the
light of cluster galaxies were added to the simulated lensed images. A
mass-matched sample of $10$ X-ray-selected clusters (Smith et al.
2005) observed at high angular resolution with the Hubble Space
Telescope (HST) was used for comparison with the simulated sample.
Finally, an automated objective arc-detection algorithm was applied to
both the observed and the simulated samples. This procedure permitted measuring and comparing the frequency of arcs
over a larger range in magnitudes (down to $R \leq 24$ mag). H05 found that the
lensing efficiency of their simulated clusters at $z\approx 0.2$ was
consistent, to within Poisson errors, with that of their observed
sample. While the analysis suggested that the observed clusters could
be somewhat more efficient lenses by up to a factor of two, this
conclusion was limited by the small size of both the observed and the
simulated samples, as well as the parameters assumed in the
simulations.

Indeed, an important parameter that affects all theoretical studies of
arc statistics is $\sigma_{8}$, the overdensity within an $8~{\rm
  Mpc}$ radius comoving sphere. Past simulations have used diverse
values: 0.9 (B98; Dalal et al. 2004; H05) or 0.95 (Wambsganss et al.
2004; Hennawi et al.  2007). Fedeli et al. (2008) have recently
analyzed the effect of $\sigma_{8}$ on the arc statistics question,
and pointed out that the most recent values of $\sigma_{8}$ from WMAP5
($0.796\pm 0.036$; Dunkley et al. 2009) revive and reinforce the
discrepancy between theory and observations of arc statistics.

A possibly related debate has emerged recently on the subject of the
size of the Einstein radius in clusters. Broadhurst and Barkana (2008)
calculated the distribution of Einstein radii in clusters with a
spherical Navarro, Frenk, \& White (NFW; 1996) profile, and with a
concentration distribution according to Neto et al. (2007). They
compared their prediction with the observed Einstein radii of three
clusters, among them Abell 1689, and found that the observed radii are
significantly larger than the theoretical expectation. Yet another
cluster with a large Einstein radius was recently reported by Zitrin
et al. (2009). Sadeh \& Rephaeli (2008) have calculated the
concentration distribution of clusters based on the distribution of
cluster formation times. They too find a discrepancy, albeit weak,
between the observed Einstein radius of Abell 1689 and its expected
value. Oguri \& Blandford (2009), however, argue that the Einstein
radius they obtain using a generalized triaxial form of the NFW
profile (Jing \& Suto 2002) is consistent with that observed in Abell
1689. In addition, they provide a prediction for the distribution of
Einstein radii, which can be tested with a large statistical cluster
sample.

Clearly, resolution of these problems requires, on the theoretical side ,
improved simulations, incorporating the most realistic cosmological
parameters, source parameters, and observational effects; and from the
observational perspective, large, well-understood samples of clusters
at various redshifts, selected by diverse methods and uniformly
observed at the high depth and resolution needed for the clear
detection of large arcs.

In this paper, we address this observational perspective. We explore
the observed statistical properties of 97 galaxy clusters imaged with
HST. This cluster sample is large enough to be separated into several
subsamples based on redshift and selection type. We apply two
different arc detection algorithms to the clusters, and compile a
high-resolution arc catalogue. We then study the arc statistics in the
various subsamples. In a forthcoming publication, we will compare the
observed statistics of this sample to new, improved, calculations of
matched simulated samples. Throughout this paper we adopt a $\Lambda
{\rm CDM}$ cosmology with parameters $\Omega_{\rm m}=0.3$,
$\Omega_{\rm \Lambda}=0.7$, and $H_{0}=70~{\rm km~s}^{-1}~{\rm
  Mpc}^{-1}$. Magnitudes are in the Vega system.

\section{Cluster Samples and Analysis}

We have compiled from the HST archive a sample of clusters
observed with the Advanced Camera for Surveys (ACS). The ACS has a
field of view $3\farcm 3 \times 3\farcm 3$, a pixel scale of $0\farcs
05$, and a point-spread function full width at half maximum of
$\approx 0\farcs 1$.

Among the clusters in our sample, $35$ are from the MAssive Cluster
Survey (MACS; Ebeling, Edge, \& Henry, 2001), and $52$ are from the
Red-Sequence Cluster Survey (RCS; Gladders \& Yee, 2005). To these we
add the $10$ clusters of Smith et al. (2005), observed with
WFPC2, and already analyzed in H05, for a total of $97$ clusters. Each
of the three WFPC2 WF CCDs had a FOV of $1\farcm 3 \times 1\farcm 3$
and a pixel scale of $0\farcs1$. We begin with a brief summary of the
relevant details of each of these surveys.

\subsection{The MAssive Cluster Survey (MACS)}

\begin{table*}
\caption{MACS low-redshift ($0.3\leq z < 0.5$) sample}
\smallskip
\begin{tabular}{lccrc}

\hline
\noalign{\smallskip}
Cluster & RA & Dec & $z$ (ref) & BCG \\
        &     & & &  $m_{F606W}$ \\ 
\noalign{\smallskip}
\hline
\noalign{\smallskip}

MACSJ0035.4$-$2015  & 00:35:26.2 & $-20:15:44.2$ & ... (1) &
19.55\\
MACSJ0916.1$-$0023  & 09:16:11.5 & $-00:23:46.6$ & ... (1) &
----\\
MACSJ0140.0$-$0555  & 01:40:00.9 & $-05:55:02.0$ & ... (1) &
20.25\\
MACSJ0140.0$-$3410  & 01:40:05.6 & $-34:10:39.7$ & ... (1) &
19.91\\
MACSJ0152.5$-$2852  & 01:52:34.4 & $-28:53:37.4$ & ... (1) &
20.32\\
MACSJ0451.9$+$0006  & 04:51:54.7 & $+00:06:17.3$ & 0.430 (2)&
20.16\\
MACSJ0520.7$-$1328  & 05:20:42.0 & $-13:28:47.6$ & ... (1) &
19.29\\ 
MACSJ0712.3$+$5931  & 07:12:20.4 & $+59:32:20.8$ & 0.328 (2)& 
19.00\\
MACSJ0845.4$+$0327  & 08:45:27.8 & $+03:27:38.8$ & ... (1) &
19.26\\
MACSJ0947.2$+$7623  & 09:47:13.2 & $+76:23:12.7$ & 0.345 (3)&
18.89\\
MACSJ0949.8$+$1708  & 09:49:51.8 & $+17:07:08.8$ & ... (1) &
19.81\\
MACSJ1006.9$+$3200  & 10:06:54.7 & $+32:01:32.3$ & ... (1) &
19.36\\ 
MACSJ1115.2$+$5320  & 11:15:14.8 & $+53:19:54.6$ & ... (1) &
19.68\\ 
MACSJ1115.8$+$0129  & 11:15:51.9 & $+01:29:54.2$ & 0.355 (3)&
19.52\\ 
MACSJ1133.2$+$5008  & 11:33:13.3 & $+50:08:39.1$ & 0.389 (4)&
19.64\\
MACSJ1206.2$-$0847  & 12:06:12.2 & $-08:48:04.4$ & 0.440 (5)&
19.92\\ 
MACSJ1236.9$+$6311  & 12:36:58.8 & $+63:11:12.2$ & 0.302 (4)&
18.91\\  
MACSJ1258.0$+$4702  & 12:58:02.1 & $+47:02:53.5$ & ... (1) &
19.64\\
MACSJ1319.9$+$7003  & 13:20:08.5 & $+70:04:39.0$ & ... (1) &
19.07\\ 
MACSJ1354.6$+$7715  & 13:54:30.6 & $+77:15:20.9$ & 0.396 (4)&
----\\
MACSJ1652.3$+$5534  & 16:52:18.8 & $+55:34:56.5$ & ... (1) &
19.27\\
MACSJ2135.2$-$0102  & 21:35:12.1 & $-01:02:57.2$ & 0.33 (6)&
19.24\\ 
MACSJ2243.3$-$0935  & 22:43:20.2 & $-09:35:26.9$ & ... (1) &
----\\ 
\noalign{\smallskip}
\hline
\smallskip

\end{tabular} 

Notes - Last column gives the magnitude of the brightest cluster
galaxy. Redshift references: (1) Ebeling et al., in preparation; (2)
Stott et al. (2007); (3) Allen et
al. (2008); (4) Edge et al. (2003); (5) Balestra et al. (2007); (6) Smail et
al. (2007).

\end{table*}

\begin{table*}

\caption{MACS medium-redshift ($0.5 \leq z < 0.7$) sample}
\smallskip
\begin{tabular}{lcccccc}
\hline
\noalign{\smallskip}
Cluster & RA & Dec & $z$ & ${\rm L}_{{\rm X}}(0.1-2.4{\rm ~keV})$  &
${\rm M}_{200}$ & BCG  \\
        & (J2000) & (J2000) & & [$10^{44}\, {\rm erg \,s}^{-1}$]  &
        [$10^{15} M_{\odot}$] & $m_{F814W}$ \\ 
\noalign{\smallskip}
\hline
\noalign{\smallskip}

MACSJ0018.5$+$1626 & 00:18:33.8  & $+16:26:16.6$ & 0.546 & 19.6 &  3.3 & 19.6\\
MACSJ0025.4$-$1222 & 00:25:29.4  & $-12:22:37.1$ & 0.584 &  8.8 &  1.8 & ----\\
MACSJ0257.1$-$2325 & 02:57:08.8  & $-23:26:03.3$ & 0.505 & 13.7 &  2.5 & 18.3\\
MACSJ0454.1$-$0300 & 04:54:11.1  & $-03:00:53.8$ & 0.538 & 16.8 &  2.9 & 18.9\\
MACSJ0647.7$+$7015 & 06:47:50.1  & $+70:14:56.4$ & 0.591 & 15.9 &  2.8 & 18.9\\
MACSJ0717.5$+$3745 & 07:17:32.9  & $+37:45:05.4$ & 0.546 & 24.6  & 3.9 &
----\\
MACSJ0744.8$+$3927 & 07:44:52.8  & $+39:27:26.7$ & 0.698 & 22.9 &  3.7 & 19.1\\
MACSJ0911.2$+$1746 & 09:11:11.2  & $+17:46:34.8$ & 0.505 &  7.8 &  1.6 & 18.8\\
MACSJ1149.5$+$2223 & 11:49:35.5  & $+22:24:04.2$ & 0.544 & 17.6 &  3.0 & 18.9\\
MACSJ1423.8$+$2404 & 14:23:48.6  & $+24:04:49.1$ & 0.543 & 16.5 &  2.9 & 18.1\\
MACSJ2129.4$-$0741 & 21:29:26.3  & $-07:41:26.2$ & 0.589 & 15.7 &  2.6 & 19.5\\
MACSJ2214.9$-$1359 & 22:14:57.3  & $-14:00:12.2$ & 0.503 & 14.1 &  2.5 & 18.2\\
        
\noalign{\smallskip}
\hline
\smallskip
\end{tabular} 

Note - Redshifts and X-ray luminosities are from Ebeling et
al. (2007). ${\rm M}_{200}$ are based on the ${\rm L}_{\rm X}-{\rm
  M}_{200}$ relation of Reiprich \& B{\" o}hringer (2002). Last column
gives the magnitude of the brightest cluster galaxy. 
\end{table*} 

MACS (Ebeling et al., 2001) has provided a statistically complete,
X-ray selected sample of the most X-ray luminous galaxy clusters at
$z>0.3$. Based on sources detected in the R\"ontgen Satellit (ROSAT)
All-Sky Survey (RASS, Voges et al. 1999), MACS covers 22,735 deg$^2$
of extragalactic sky ($|b|>20$ deg); the present MACS sample,
estimated to be at least 90\% complete, comprises 124 clusters all of
which have spectroscopic redshifts. Owing to the high X-ray flux limit
of the RASS and the lower redshift limit of $z=0.3$, MACS clusters
feature X-ray luminosities of, typically, 5--10$\times 10^{44}$ erg
s$^{-1}$ in the 0.1--2.4 keV band (Ebeling et al.\ 2007). MACS thus
probes the high end of the cluster mass function, including some of
the most powerful gravitational lenses (Smith et al.\ 2009; Zitrin et
al.\ 2009); see also Smail et al. (2007) for a spectacular case of
galaxy-galaxy lensing in the field of a MACS cluster. MACS clusters
 have been used for a wide range of cosmological and astrophysical
applications, e.g., in cosmological studies (Allen
et al. 2008; Mantz et al. 2008, 2009a, b), investigations of
large-scale structure (Ebeling, Barrett, \&
Donovan\ 2004; Kartaltepe et al.\ 2008), and
 studies of the galaxy content and gas properties 
of individual clusters (e.g., Ma et al.
2008, 2009).

Here we use 
images of 35 MACS clusters observed with the ACS (GO-09722,
GO-10491, GO-10875, PI Ebeling).

We divide these clusters into two subsamples according to redshift,
$0.3 \leq z < 0.5$, and $0.5 \leq z < 0.7$, which consist of 23 and 12
clusters, respectively. The low-redshift sample was observed with {\it
  HST} in Snapshot mode, meaning the telescope schedulers chose a fraction
of the targets from the full MACS sample, based solely on their
scheduling convenience. Thus, the
clusters we analyse are
 an unbiased, representative
selection from the entire MACS sample. The medium-redshift sample consists of a complete
set of 12 MACS clusters in this redshift range that are visible from Hawaii. Strong-lensing mass reconstructions of the clusters in this subsample have been recently presented by Zitrin et al. (2010).

The low-redshift clusters were observed
through the ${\rm F}606{\rm W}$ filter (mean wavelength $\sim
6060~{\rm \AA}$) with exposure times of $1200$ s, while the
medium-redshift sample was observed through the ${\rm F}814{\rm W}$
filter (mean wavelength $\sim 8140~{\rm \AA}$) with exposure times of
$\sim 4500$ s. Applying the ${\rm
  L_{X}-M_{200}}$ relation\footnote{${\rm M}_{200}$ is the mass
  enclosed within $r_{200}$, the radius within which the average
  density is equal to $200$ times the critical cosmological density at
  the observed redshift.} of Reiprich \& B{\" o}hringer (2002) yields a cluster mass range of $(1.4 \leq
{\rm M}_{200} \leq 4.1)\times 10^{15}{\rm M}_{\odot}$, and $(1.6 \leq
{\rm M}_{200} \leq 3.9)\times 10^{15}{\rm M}_{\odot}$ for the low- and
medium-redshift samples, respectively. The cluster properties are
listed in Tables 1 and 2.

\subsection{X-ray Brightest Abell-type Clusters of
  galaxies (XBACs) $z\approx 0.2$ sample}

The sample of Smith et al. (2005), as analyzed in H05, consists of 10
galaxy clusters from the X-ray Brightest Abell-type Clusters of
galaxies (XBACs) catalogue (Ebeling et al. 1996), with $0.17< z <0.26$.
The $0.1-2.4$ keV flux limit of $f_{X}\geq 5.0\times10^{-12}$ erg
cm$^{-2}$ s$^{-1}$ applied to this redshift range implies X-ray
luminosities $L_{X}\geq4.1\times10^{44}$ erg s$^{-1}$, i.e.
similar to the MACS clusters at their higher redshifts. Details of this
sample and its properties can be found in table 1 of H05.

\subsection{The Red-Sequence Cluster Survey (RCS)}

The RCS survey was conducted using the Canada-France-Hawaii Telescope
(CFHT) through the $R_{c}$ and $z'$ filters. Gladders \& Yee (2005)
applied a red-sequencing technique to an area of $\sim 100 {\rm
  ~deg}^{2}$, and a catalogue of $\sim 1000$ clusters at $0.2<z<1.4$
was compiled. The survey is complete to $5\sigma$ magnitude limits of
$24.9$ and $23.8$ in $z'$ and $R_{c}$, respectively. Like MACS clusters, RCS clusters have also been used in many applications, e.g., studying the scaling relations between different cluster properties (Hicks et al. 2008), and exploring the evolution of the red-sequence galaxy luminosity function (Gilbank et al. 2008).

Among the RCS
clusters, a subset of $150$ clusters was proposed for HST observation,
again in Snapshot mode, out of which $52$ were selected by HST
schedulers based on scheduling conveneience, and imaged using ACS,
(GO-10626, PI Loh). Contrary to the MACS and XBACs clusters that we
analyse here, which were
chosen in an unbiased way from among complete samples, we do not know
what were the criteria, if any, for selecting the 150 RCS clusters to be
potential HST Snapshot targets. We suspect that there may have been
some bias toward including clusters that already had evidence of
strong lensing, based on previous ground-based imaging. However, it is
highly unlikely that the 150 clusters were chosen, intentionally or
unintentionally, in a way that would {\it avoid} systems with strong
lensing (and it is also hard to imagine a logical reason for such a
choice). 
The main result of our study will be that the
RCS
clusters observed by HST
 are inefficient as lenses, when compared to the truly unbiased sample of X-ray selected
clusters. This conclusion, applied to the
RCS clusters as a whole, will therefore only be
strengthened, if the 150 RCS clusters were pre-selected to favor strong
lenses. Our results will thus provide a firm
and useful upper limit on the RCS lensing fraction.

The
clusters were imaged through the ${\rm F}814{\rm W}$ filter with
exposure times of $1440$~s. Luminosities and mass estimates 
of the RCS clusters have not
been published to date. In $\S 4$ below, we show that the RCS
clusters and the X-ray selected clusters above have similar optical
luminosities.

As with the X-ray selected clusters above, we divide the RCS clusters
into redshift bins: the same low ($0.3 \leq z < 0.5$) and medium ($0.5
\leq z < 0.7$) redshift subsamples which were defined above, and a
third, high-redshift, subsample at $0.7 \leq z \leq 1$. The three
redshift subsamples consist of $18$, $18$, and $16$ clusters,
respectively. The properties of the $52$ RCS clusters are listed in
Tables 3, 4, and 5.
\begin{table}
\caption{RCS low-redshift ($0.3 \leq z < 0.5$) sample}
\smallskip
\begin{tabular}{lcccc}
\hline
\noalign{\smallskip}
Cluster & RA & Dec & $z$ & BCG  \\
        &    &     &     & $m_{F814W}$ \\   
\noalign{\smallskip}
\hline
\noalign{\smallskip}
RCS022403$-$0227.7    &    02:24:03.4&   $-02:27:52.1$ &  0.314 & ----\\
RCS035139$-$0956.4    &    03:51:39.5&   $-09:56:32.6$ &  0.334 & 17.3\\
RCS044406$-$2820.5    &   04:44:06.4&    $-28:20:37.9$ &  0.437 & 18.0\\
RCS051536$-$4325.5    &   05:15:37.0&    $-43:25:31.1$ &  0.44 & 18.3\\
RCS051834$-$4325.1    &   05:18:35.2&    $-43:25:15.0$ &  0.475 & 18.6\\
RCS092821$+$3646.5    &   09:28:22.3&    $+36:46:31.9$ &  0.356 & 18.2\\
RCS110233$-$0319.2    &   11:02:33.5&    $-03:19:19.3$ &  0.423 & 17.6\\
RCS110258$-$0521.2    &   11:02:59.2&    $-05:21:13.9$ &  0.395 & 18.4\\
RCS110340$-$0458.1    &   11:03:40.7&    $-04:58:12.0$ &  0.492 & 19.3\\
RCS131912$-$0206.9    &   13:19:12.7&    $-02:06:59.7$ &  0.354 & ----\\
RCS145226$+$0834.6    &   14:52:27.3&    $+08:34:36.7$ &  0.325 & 18.1\\
RCS145900.4$+$102336    & 14:59:00.8&    $+10:23:34.5$ &  0.395 & 18.4\\
RCS151110.7$+$100203    & 15:11:11.1&    $+10:02:05.9$ &  0.455 & 18.1\\
RCS151306.9$+$061124    & 15:13:06.5&    $+06:11:24.8$ &  0.325 & 16.4\\
RCS211519$-$6309.5    &   21:15:20.3&    $-63:09:31.0$ &  0.331 & 17.6\\
RCS212134$-$6335.8    &   21:21:35.0&    $-63:35:50.9$ &  0.351 & 17.2\\
RCS215609.1$+$012319    & 21:56:09.3&    $+01:23:23.1$ &  0.335 & 16.9\\
RCS223952$-$6044.8    &   22:39:52.8&    $-60:44:53.6$ &  0.429 & 18.7\\
\noalign{\smallskip}
\hline
\smallskip
\end{tabular} 
\end{table}
\begin{table}
\caption{RCS medium-redshift ($0.5 \leq z < 0.7$) sample}
\smallskip
\begin{tabular}{lcccc}
\hline
\noalign{\smallskip}
Cluster & RA & Dec & $z$ & BCG  \\
        &    &     &     & $m_{F814W}$ \\
\noalign{\smallskip}
\hline
\noalign{\smallskip}
RCS033414$-$2824.6    &    03:34:14.5&   $-28:24:34.5$ &  0.668 & ----\\
RCS035027$-$0855.1    &    03:50:27.4&   $-08:55:13.5$ &  0.584 & 19.0\\  
RCS044207$-$2815.0    &   04:42:08.1&    $-28:15:11.3$ &  0.522 & 18.9\\
RCS051128$-$4235.2    &   05:11:27.8&    $-42:35:11.6$ &  0.518 & 18.3\\
RCS051855$-$4315.0    &   05:18:55.0&    $-43:15:00.9$ &  0.544 & 19.0\\
RCS051919$-$4247.8    &   05:19:19.8&    $-42:47:49.2$ &  0.603 & 19.9\\ 
RCS110104$-$0351.3    &   11:01:04.7&    $-03:51:21.3$ &  0.639 & ----\\  
RCS110733$-$0520.6    &   11:07:33.5&    $-05:20:39.4$ &  0.597 & 18.7\\
RCS110752$-$0516.5    &   11:07:53.0&    $-05:16:35.0$ &  0.579 & 19.4\\
RCS110814$-$0430.8    &   11:08:14.5&    $-04:30:53.9$ &  0.638 & ----\\
RCS131722$-$0201.4    &   13:17:22.8&    $-02:01:28.8$ &  0.535 & 18.2\\
RCS132335$+$3022.6    &   13:23:35.5&    $+30:22:43.7$ &  0.538 & 18.1\\
RCS141910$+$5326.1    &   14:19:10.3&    $+53:26:07.5$ &  0.647 & 19.4\\
RCS151840.1$+$084500    & 15:18:40.3&    $+08:45:05.0$ &  0.515 & 18.7\\
RCS161547$+$3057.3    &   16:15:47.5&    $+30:57:14.1$ &  0.514 & 18.8\\
RCS215223$-$0503.8    &   21:52:23.2&    $-05:03:44.2$ &  0.545 & 18.7\\
RCS231654$-$0011.1    &   23:16:54.8&    $-00:11:06.8$ &  0.56 & 19.6\\
RCS234717$-$3634.4    &   23:47:17.4&    $-36:34:32.6$ &  0.537 & 18.8\\
\noalign{\smallskip}
\hline
\smallskip
\end{tabular} 
\end{table}
\begin{table}
\caption{RCS high-redshift ($0.7 \leq z \leq 1$) sample}
\smallskip
\begin{tabular}{lccc}
\hline
\noalign{\smallskip}
Cluster & RA & Dec & $z$ \\
\noalign{\smallskip}
\hline
\noalign{\smallskip}
RCS022453$-$0316.7    &    02:24:53.6&   $-03:16:47.5$ &  0.906 \\  
RCS025242.5$-$150024  &    02:52:42.7&   $-15:00:28.0$ &  0.995 \\  
RCS043934$-$2904.6    &   04:39:34.2&    $-29:04:43.9$ &  0.786 \\
RCS051940$-$4402.1    &   05:19:40.3&    $-44:02:13.8$ &  0.913 \\
RCS110439$-$0445.0    &   11:04:40.3&    $-04:45:03.2$ &  0.715 \\
RCS110651$-$0350.3    &   11:06:52.2&    $-03:50:23.8$ &  0.768 \\
RCS110723$-$0523.2    &   11:07:23.8&    $-05:23:16.1$ &  0.794 \\
RCS112225$+$2422.9    &   11:22:25.5&    $+24:22:51.3$ &  0.799 \\
RCS132939$+$2853.3    &   13:29:39.8&    $+28:53:14.3$ &  0.901 \\
RCS145039$+$0840.7    &   14:50:40.2&    $+08:40:46.9$ &  0.769 \\
RCS162009$+$2929.4    &   16:20:09.2&    $+29:29:33.8$ &  0.797 \\
RCS211852$-$6334.6    &   21:18:52.6&    $-63:34:43.1$ &  0.786 \\
RCS212238$-$6146.1    &   21:22:38.3&    $-61:46:17.0$ &  0.856 \\
RCS215248$-$0609.4    &   21:52:49.2&    $-06:09:24.4$ &  0.704 \\
RCS231831$+$0034.3    &   23:18:31.8&    $+00:34:22.8$ &  0.809 \\
RCS234220$-$3534.3    &   23:42:20.4&    $-35:34:15.5$ &  0.802 \\
\noalign{\smallskip}
\hline
\smallskip
\end{tabular} 
\end{table}

\subsection{Arc detection}

In H05, we introduced the use of an automated arc detection algorithm 
to arc
statistics studies. Automated arc detection is important for an
objective, quantitative, and fair comparison of arc statistics in
observed and simulated data.  In the meantime, a number of other
arc-detection algorithms have been published, by Lenzen et al. (2004),
Alard (2006), and Seidel \& Bartelmann (2007; SB07). In the present
work, we subject all of the images to two of these algorithms, H05 and
SB07.

The H05 arc-detection algorithm is based
on application of the SExtractor (Bertin \& Arnouts 1996) object
identification software. The output of repeated SExtractor calls, using
different detection parameters each time, is
filtered using some threshold of object elongation. The final
SExtractor call is executed on an image combined from the filtered
``segmentation image'' outputs of the previous calls.  The arc
candidates detected in that last call are included in the final arc
catalogue if they meet the required detection parameters defined by the
user.

The SB07 algorithm is based on light moments. The image is divided
into small cells which are iteratively moved to their local light
centres. Then, for each cell, an ellipticity vector is calculated
using light moments. Adjacent cells with similarly oriented
ellipticity vectors are joined together and considered as part of an
arc candidate, whose outer boundary is determined by an active contour
method. Candidates are accepted if they conform to specified
parameters.

In the present work, we apply an acceptance criterion on arc
length-to-width ratio of $l/w \geq 8$. We also use a magnitude limit
of $m \leq 24$ as another acceptance criterion which, given the
exposure times of our sample, results in the detection of arcs with
signal-to-noise $S/N \ga 3$. Our magnitude limit is higher than most of the magnitude limits used in previous studies,  such as B98 and Zaritsky \& Gonzales 2003, allowing us to include fainter arcs in our analysis. 
Nevertheless, our acceptance threshold for arc detections is brighter than the arc detection limits of all the
images, with their range of exposure times and filters, thus permitting a
meaningful comparison of arc statistics among the various subsamples. 
This holds also
for the WFPC2 images of the XBACS sample. Although WFPC2 was less
sensitive than ACS, the WFPC2 exposure times were longer, typically
7000~s, leading to similar depths. Furthermore, the somewhat lower angular
resolution of WFPC2, due to its larger pixels ($0\farcs 1$), is not
important, since the arcs we consider are always much larger, and all the
arcs we find below in ACS images would have been detected in long
WFPC2 exposures as well.
We note that we use total-magnitude limit for arcs, rather than
considering surface brightness, which could also plausibly be used.
We do this to conform with previous observational and theoretical studies,
but also because arcs, especially at HST resolution, display rich
structure and unresolved clumps, and hence it is not clear that
mean surface brightness would be a more relevant observable. 
Due to the varying position of the cluster
centres within the FOV, the cluster coverage area varies.  We
therefore also limit our search to a $60''$ radius from the cluster
centre. The automated arc detection results were visually inspected in
order to remove false positives such as spikes from saturated stars,
galaxy spiral arms, and edge-on galaxies.

While most of the arcs in
our sample are detected by both programs, a few unmistakable lensed
arcs are picked out by only one or the other. The SB07 arcfinder is
more 
sucessful than the H05 arcfinder in detecting arcs that are
superimposed on the light of cluster galaxies. 
On the other hand, the H05 arcfinder produce a better ``segmentation''
compared to the SB07 arcfinder, which sometimes breaks 
arcs into smaller arclets, which then do not qualify as giant arcs. 
We defer a more detailed comparison of these and
 other arcfinders to a future study. 

\section{Results}

Figure 1 shows the ACS images of the clusters in which arcs are
detected, and Figure 2 provides zoom-ins on the individual arc features.
Table 6 lists the properties of the detected arcs, which we discuss in
more detail below.
\begin{figure*}
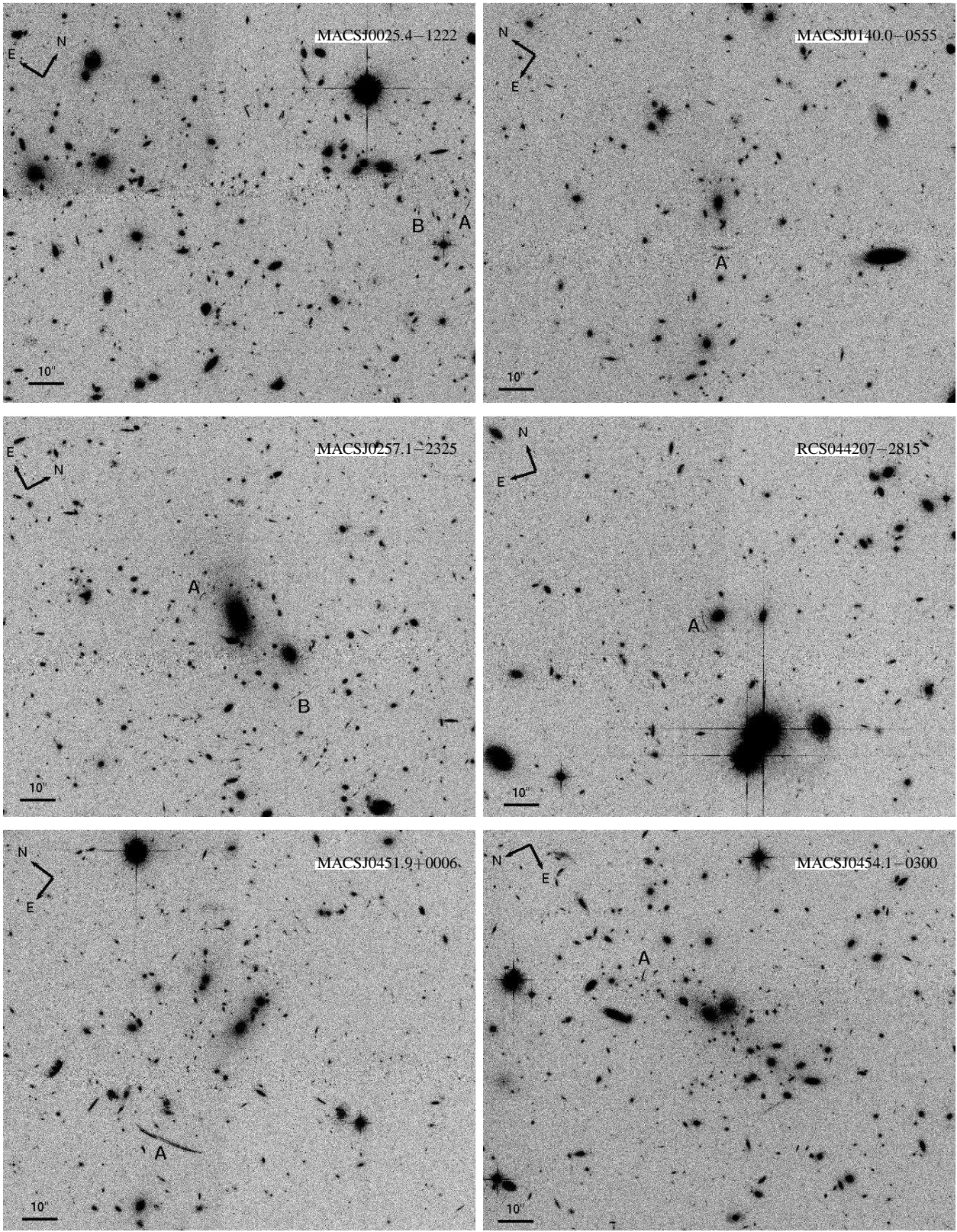

\centering
\footnotesize

\makebox{
\begin{lpic}{51020_arcs(0.42)}
\setlength{\lpbgboxsep}{0.2mm}
\lbl[tlW]{130,155;MACSJ0025.4$-$1222}
\end{lpic}

\begin{lpic}{j9fc13010_arcs2(0.42)}
\setlength{\lpbgboxsep}{0.2mm}
\lbl[tlW]{130,155;MACSJ0140.0$-$0555}
\end{lpic}
}

\vskip 0.2cm
\makebox{
\begin{lpic}{j8qu02020_arcs2(0.42)}
\setlength{\lpbgboxsep}{0.2mm}
\lbl[tlW]{130,155;MACSJ0257.1$-$2325}
\end{lpic}

\begin{lpic}{j9en23010_arcs2(0.42)}
\setlength{\lpbgboxsep}{0.2mm}
\lbl[tlW]{130,155;RCS044207$-$2815}
\end{lpic}
}

\vskip 0.2cm
\makebox{
\begin{lpic}{j9fc30010_arcs2(0.42)}
\setlength{\lpbgboxsep}{0.2mm}
\lbl[tlW]{130,155;MACSJ0451.9$+$0006}
\end{lpic}

\begin{lpic}{14010_arcs_A(0.42)}
\setlength{\lpbgboxsep}{0.2mm}
\lbl[tlW]{130,155;MACSJ0454.1$-$0300}
\end{lpic}
}

\normalsize

\caption{$2\farcm2 \times 1\farcm9$ sections of the HST/ACS images of
  the clusters, showing the location of detected arcs. See Fig. 2 for
  a detailed view of each arc.}

\end{figure*}
\addtocounter{figure}{-1}
\begin{figure*}
\centering
\footnotesize

\makebox{
\begin{lpic}{j9fc32010_arcs2(0.42)}
\setlength{\lpbgboxsep}{0.2mm}
\lbl[tlW]{130,155;MACSJ0520.7$-$1328}
\end{lpic}

\begin{lpic}{j9fc37010_arcs2(0.42)}
\setlength{\lpbgboxsep}{0.2mm}
\lbl[tlW]{130,155;MACSJ0712.3$+$5931}
\end{lpic}
}

\vskip 0.2cm
\makebox{
\begin{lpic}{j8qu05020_arcs2(0.42)}
\setlength{\lpbgboxsep}{0.2mm}
\lbl[tlW]{130,155;MACSJ0717.5$+$3745}
\end{lpic}

\begin{lpic}{j8qu06020_arcs2(0.42)}
\setlength{\lpbgboxsep}{0.2mm}
\lbl[tlW]{130,155;MACSJ0744.8$+$3927}
\end{lpic}
}

\vskip 0.2cm
\makebox{
\begin{lpic}{j9fc41010_arcs2(0.42)}
\setlength{\lpbgboxsep}{0.2mm}
\lbl[tlW]{130,155;MACSJ0947.2$+$7623}
\end{lpic}

\begin{lpic}{j9fc42010_arcs2(0.42)}
\setlength{\lpbgboxsep}{0.2mm}
\lbl[tlW]{130,155;MACSJ0949.8$+$1708}
\end{lpic}
}

\caption{[continued]}
\end{figure*}
\addtocounter{figure}{-1}
\begin{figure*}
\footnotesize

\makebox{
\begin{lpic}{j9fc46010_arcs2(0.42)}
\setlength{\lpbgboxsep}{0.2mm}
\lbl[tlW]{130,155;MACSJ1115.2$+$5320}
\end{lpic}

\begin{lpic}{j9fc47010_arcs2(0.42)}
\setlength{\lpbgboxsep}{0.2mm}
\lbl[tlW]{130,155;MACSJ1115.8$+$0129}
\end{lpic}
}

\vskip 0.2cm
\makebox{
\begin{lpic}{j9fc49010_arcs2(0.42)}
\setlength{\lpbgboxsep}{0.2mm}
\lbl[tlW]{130,155;MACSJ1133.2$+$5008}
\end{lpic}

\begin{lpic}{j8qu08020_arcs2(0.42)}
\setlength{\lpbgboxsep}{0.2mm}
\lbl[tlW]{130,155;MACSJ1149.5$+$2223}
\end{lpic}
}

\vskip 0.2cm
\makebox{
\begin{lpic}{j9fc52010_arcs2(0.42)}
\setlength{\lpbgboxsep}{0.2mm}
\lbl[tlW]{130,155;MACSJ1206.2$-$0847}
\end{lpic}

\begin{lpic}{j9fc55010_arcs2(0.42)}
\setlength{\lpbgboxsep}{0.2mm}
\lbl[tlW]{130,155;MACSJ1236.9$+$6311}
\end{lpic}
}

\caption{[continued]}
\end{figure*}
\addtocounter{figure}{-1}
\begin{figure*}
\footnotesize

\makebox{
\begin{lpic}{j9fc62010_arcs2(0.42)}
\setlength{\lpbgboxsep}{0.2mm}
\lbl[tlW]{130,155;MACSJ1354.6$+$7715}
\end{lpic}

\begin{lpic}{j9en1q010_arcs2(0.42)} 
\setlength{\lpbgboxsep}{0.2mm}
\lbl[tlW]{130,155;RCS131722$-$0201.4}
\end{lpic}
}

\vskip 0.2cm
\makebox{
\begin{lpic}{j9en1h010_arcs2(0.42)}
\setlength{\lpbgboxsep}{0.2mm}
\lbl[tlW]{130,155;RCS141910$+$5326.1}
\end{lpic}

\begin{lpic}{j8qu09020_arcs2(0.42)}
\setlength{\lpbgboxsep}{0.2mm}
\lbl[tlW]{130,155;MACSJ1423.8$+$2404}
\end{lpic}
}

\vskip 0.2cm
\makebox{
\begin{lpic}{j9en22010_arcs2(0.42)}
\setlength{\lpbgboxsep}{0.2mm}
\lbl[tlW]{130,155;RCS151840.1$+$084500}
\end{lpic}

\begin{lpic}{drz0s010_arcs2(0.42)}
\setlength{\lpbgboxsep}{0.2mm}
\lbl[tlW]{130,155;RCS212134$-$6335.8}
\end{lpic}
}

\caption{[continued]}
\end{figure*}
\addtocounter{figure}{-1}
\begin{figure*}
\footnotesize

\makebox{
\begin{lpic}{j8qu10020_arcs2(0.42)}
\setlength{\lpbgboxsep}{0.2mm}
\lbl[tlW]{130,155;MACSJ2129.4$-$0741}
\end{lpic}

\begin{lpic}{drz0p010_arcs2(0.42)}
\setlength{\lpbgboxsep}{0.2mm}
\lbl[tlW]{130,155;RCS215609.1$+$012319}
\end{lpic}
}

\vskip 0.2cm
\begin{lpic}{j8qu11020_arcs2(0.42)}
\setlength{\lpbgboxsep}{0.2mm}
\lbl[tlW]{130,155;MACSJ2214.9$-$1359}
\end{lpic}

\caption{[continued]}
\end{figure*}

\begin{table*}
\caption{Detected arcs and properties}
\smallskip
\begin{tabular}{lcccccc}
\hline
\noalign{\smallskip}
Cluster & Arc ID & Length & Width & $l/w$ & $m_{\rm F606W}$/ &
Radial separation \\
 & & & & & $m_{\rm F814W}$ & from cluster centre \\
 & & [arcsec] & [arcsec] & & [mag] & [arcsec] \\ 
\noalign{\smallskip}
\hline
\noalign{\smallskip}

MACSJ0025.4$-$1222   & A  &  3.3 & 0.2 &  13.9 & 23.6 & ---- \\
                     & B  &  2.7 & 0.3 & 9.64 & 23.7 & ---- \\  
MACSJ0140.0$-$0555   & A  &  5.1 & 0.6 &  8.3 & 21.8 & 13.7 \\
MACSJ0257.1$-$2325     & A  &  3.0 & 0.3 & 10.6 & 23.7 & 11.6 \\
                   & B  &  3.5 & 2.8 & 12.7 & 23.5 & 27.8 \\
RCS044207$-$2815.0   & A  &  4.3 & 0.2 & 18.5 & 23.0 &  4.5 \\
MACSJ0451.9$+$0006   & A  & 20.2 & 0.7 & 29.3 & 20.5 & 38.3 \\
MACSJ0454.1$-$0300   & A  & 3.7  & 0.4 & 10.4 & 22.3 & 21.8 \\
MACSJ0520.7$-$1328   & A  &  6.1 & 0.5 & 12.2 & 22.2 & 30.9 \\ 
                   & B1 &  2.3 & 0.3 &  8.6  & 23.7 & 24.1 \\
                   & B2 &  2.6 & 0.3 &  8.8 & 23.5 & 27.4 \\
MACSJ0712.3$+$5931   & A  &  5.2 & 0.3 & 17.9 & 22.8 & 23.2 \\
MACSJ0717.5$+$3745     & A  &  4.2 & 0.3 & 15.4 & 23.0 & ---- \\
                   & B  &  3.0 & 0.4 &  8.5 & 22.7 & ---- \\
                   & C  &  4.6 & 0.2 & 19.0 & 23.1 & ---- \\
MACSJ0744.8$+$3927     & A  &  6.1 & 0.5 & 12.3  & 20.5 & 23.3 \\
                   & B  &  5.1 & 0.2 & 27.3  & 23.9 & 35.0 \\
MACSJ0947.2$+$7623   & A  &  6.4 & 0.4 & 14.3 & 22.3 & 13.5 \\
                   & B  &  2.9 & 0.3 & 14.7 & 22.6 & 41.7 \\
                   & C  &  6.5 & 0.3 & 23.0 & 23.4 & 19.1 \\
MACSJ0949.8$+$1708   & A  &  3.3 & 0.4 &  8.8 & 22.8 & 37.6 \\
MACSJ1115.2$+$5320   & A  &  3.2 & 0.3 & 10.2 & 23.6 & 31.6 \\ 
                   & B1 &  3.4 & 0.3 & 10.3 & 22.9 & 56.2 \\
                   & B2 &  3.5 & 0.4 &  8.5 & 23.5 & 57.3 \\
                   & C  &  4.4 & 0.5 &  8.5 & 22.3 & 34.5 \\
MACSJ1115.8$+$0129   & A$^1$  &  4.8 & 0.3 & 15.1 & 23.2 & 11.2 \\
                   & B  &  5.1 & 0.2 & 26.7 & 23.7 & 37.2 \\ 
MACSJ1133.2$+$5008   & A  & 11.7 & 0.8 & 14.6 & 21.0 & 10.5 \\
                   & B  &  2.5 & 0.2 & 10.2 & 23.5 & 17.0 \\
MACSJ1149.5$+$2223     & A  &  4.0 & 0.5 &  8.4 & 22.2 & 26.0 \\
MACSJ1206.2$-$0847   & A$^2$  & 13.9 & 0.5 & 26.8 & 21.1 & 20.7 \\
                   & B$^2$  &  4.8 & 0.5 &  8.8 & 22.4 & 23.8 \\
                   & C  &  3.7 & 0.5 &  8.2 & 23.3 & 59.3 \\
MACSJ1236.9$+$6311   & A  &  2.9 & 0.3 & 10.6 & 23.8 & 38.0 \\ 
MACSJ1354.6$+$7715   & A  &  7.7 & 0.6 & 12.9 & 21.7 & ---- \\
                   & B  &  7.6 & 0.9 &  8.8 & 21.1 & ---- \\
                   & C  &  3.4 & 0.2 & 14.6 & 23.8 & ---- \\
                   & D  &  4.6 & 0.3 & 17.6 & 23.4 & ---- \\
RCS131722$-$0201.4   & A  &  2.6 & 0.3 & 10.1 & 23.0 & 48.4 \\
RCS141910$+$5326.1   & A$^3$  & 10.5 & 0.6 & 17.3 & 20.2 & 10.0 \\
                   & B$^3$  &  3.8 & 0.3 & 11.3 & 22.4 & 17.2 \\
MACSJ1423.8$+$2404     & A  &  4.2 & 0.4 & 10.3 & 22.7 & 19.7 \\
                   & B  &  3.6 & 0.2 & 14.5  & 23.3 & 20.6 \\
RCS151840.1$+$084500 & A  &  3.8 & 0.3 & 11.9 & 21.6 & 11.2 \\
RCS212134$-$6335.8   & A  &  3.8 & 0.4 & 10.4 & 21.6 &  3.4 \\  
MACSJ2129.4$-$0741     & A  &  5.4 & 0.3 & 17.3 & 22.1 & 31.5 \\
RCS215609.1$+$012319 & A  &  3.7 & 0.3 & 11.2 & 22.8 & 18.3 \\
MACSJ2214.9$-$1359     & A  &  4.0 & 0.3 & 12.4  & 23.3 & 14.3 \\
                   & B  &  4.9 & 0.3 & 17.9  & 23.0 & 20.1 \\

\noalign{\smallskip}
\hline
\smallskip
\end{tabular} 

Notes: Arcs previously reported by: $^1$Sand et al. (2005); $^2$Ebeling et al. (2009); $^3$Gladders et al. (2003).
 
\end{table*}

\subsection{X-ray selected clusters}

In the MACS sample we identify a total of $26$ arcs in $12$ out of the
$23$ low-redshift clusters, and a total of $16$ arcs in $9$ out of the
$12$ medium-redshift clusters. All but 3 of these arcs (in two
clusters) have not been previously reported (see Table 6). 
The arcs span a magnitude range of $20 <
m < 24$ and a $l/w$ ratio range of $8-29$. As shown in Figure 3, over
half of the cluster lenses, in both the low- and medium-redshift
MACS subsamples, produce multiple arcs. A similar result was found in
the XBACs sample of H05, in which 17 arcs (with $l/w \geq 8$), in 7 out
of the 10 clusters at $z\approx 0.2$, were detected.

In terms of the distribution of the angular separation of arcs from the
cluster centres, in the low-redshift MACS subsample, as shown in Fig.
4, the lensed arcs are uniformly distributed at separation angles of $10''-50''$.
In the medium-redshift MACS sample, the arcs are distributed slightly
closer to the cluster centres, but both distributions are consistent,
given the small numbers per bin. There are no arcs in this sample
beyond $35''$. Since there is an uncertainty concerning the centre
position of
the cluster MACSJ1354.6$+$7715, as discussed below, we do not include
its arcs in the above analysis.  In addition, each of the apparently
merging arc pairs MACSJ0520.7$-$1328 B1/B2 and MACSJ1115.2$+$5320 B1/B2,
are treated as one arc. We also exclude the arcs in MACSJ0717.5$+$3745
from this analysis, since this cluster is highly disturbed (Ma,
Ebeling, \& Barrett; 2009) and therefore its centre cannot be easily
determined.

\subsection{Optically selected clusters}

Only two arcs are detected in the low-redshift RCS cluster subsample.
While both arcs have $l/w \geq 10$, they are still relatively short
($<5''$) compared to some of the arcs found in the MACS sample, which
can be as long as $20''$. In the medium-redshift optical subsample,
$5$ arcs are found in $4$ out of the $18$ clusters.  
two of these arcs (in one cluster, see Table~6) have been previously reported.
No arcs are
detected among the $16$ clusters of the high-redshift ($0.7 \leq z
\leq 1$) optical subsample.

As seen in Fig.~4 and Table 6, compared to the X-ray sample, 
the arcs in the RCS sample occur at significantly
smaller separations, generally $<20''$, and sometimes only $3-5''$.
The only exception is RCS131722$-$0201.4, whose arc appears $48''$
from the cluster centre. However, as seen in Figs. 1 and 2, this arc
may actually be a small-separation image produced by the local mass
concentration traced by the galaxies near the arc. Since arcs occur
near critical curves, the small separations suggest significantly smaller
Einstein radii, and hence masses, for the RCS clusters.

\begin{figure*}

\makebox{
\footnotesize
\begin{lpic}{51020_A(0.195)}
\setlength{\lpbgboxsep}{0.2mm}
\lbl[tlW]{5,165;MACSJ0025.4$-$1222 A}
\end{lpic}

\begin{lpic}{51020_B(0.195)}
\setlength{\lpbgboxsep}{0.2mm}
\lbl[tlW]{5,165;MACSJ00025.4$-$1222 B}
\end{lpic}

\begin{lpic}{A1_o(0.195)}
\setlength{\lpbgboxsep}{0.2mm}
\lbl[tlW]{5,165;MACSJ0140.0$-$0555 A}
\end{lpic}

\begin{lpic}{02020_A(0.195)}
\setlength{\lpbgboxsep}{0.2mm}
\lbl[tlW]{5,165;MACSJ0257.1$-$2325 A}
\end{lpic}

}

\makebox{
\footnotesize

\begin{lpic}{02020_B(0.195)}
\setlength{\lpbgboxsep}{0.2mm}
\lbl[tlW]{5,165;MACSJ0257.1$-$2325 B}
\end{lpic}

\begin{lpic}{j9en23010_A(0.195)}
\setlength{\lpbgboxsep}{0.2mm}
\lbl[tlW]{5,165;RCS044207$-$2815 A}
\end{lpic}

\begin{lpic}{B1_o(0.195)}
\setlength{\lpbgboxsep}{0.2mm}
\lbl[tlW]{5,165;MACSJ0451.9$+$0006 A}
\end{lpic}

\begin{lpic}{14010_A(0.195)}
\setlength{\lpbgboxsep}{0.2mm}
\lbl[tlW]{5,165;MACSJ0454.1$-$0300 A}
\end{lpic}

}

\makebox{
\footnotesize

\begin{lpic}{C1_o(0.195)}
\setlength{\lpbgboxsep}{0.2mm}
\lbl[tlW]{5,165;MACSJ0520.7$-$1328 A}
\end{lpic}

\begin{lpic}{C2_o(0.195)}
\setlength{\lpbgboxsep}{0.2mm}
\lbl[tlW]{5,165;MACSJ0520.7$-$1328 B}
\end{lpic}

\begin{lpic}{D1_o(0.195)}
\setlength{\lpbgboxsep}{0.2mm}
\lbl[tlW]{5,165;MACSJ0712.3$+$5931 A}
\end{lpic}

\begin{lpic}{05020_A(0.195)}
\setlength{\lpbgboxsep}{0.2mm}
\lbl[tlW]{5,165;MACSJ0717.5$+$3745 A}
\end{lpic}

}

\makebox{
\footnotesize

\begin{lpic}{05020_B(0.195)}
\setlength{\lpbgboxsep}{0.2mm}
\lbl[tlW]{5,165;MACSJ0717.5$+$3745 B}
\end{lpic}

\begin{lpic}{05020_C(0.195)}
\setlength{\lpbgboxsep}{0.2mm}
\lbl[tlW]{5,165;MACSJ0717.5$+$3745 C}
\end{lpic}

\begin{lpic}{06020_A(0.195)}
\setlength{\lpbgboxsep}{0.2mm}
\lbl[tlW]{5,165;MACSJ0744.8$+$3927 A}
\end{lpic}

\begin{lpic}{06020_B(0.195)}
\setlength{\lpbgboxsep}{0.2mm}
\lbl[tlW]{5,165;MACSJ0744.8$+$3927 B}
\end{lpic}

}

\makebox{
\footnotesize

\begin{lpic}{E1_o(0.195)}
\setlength{\lpbgboxsep}{0.2mm}
\lbl[tlW]{5,165;MACSJ0947.2$+$7623 A}
\end{lpic}

\begin{lpic}{E2_o(0.195)}
\setlength{\lpbgboxsep}{0.2mm}
\lbl[tlW]{5,165;MACSJ0947.2$+$7623 C}
\end{lpic}

\begin{lpic}{j9fc41010_C(0.195)}
\setlength{\lpbgboxsep}{0.2mm}
\lbl[tlW]{5,165;MACSJ0947.2$+$7623 B}
\end{lpic}

\begin{lpic}{F1_o(0.195)}
\setlength{\lpbgboxsep}{0.2mm}
\lbl[tlW]{5,165;MACSJ0949.8$+$1708 A}
\end{lpic}

}

\makebox{
\footnotesize

\begin{lpic}{G1_o(0.195)}
\setlength{\lpbgboxsep}{0.2mm}
\lbl[tlW]{5,165;MACSJ1115.2$+$5320 A}
\end{lpic}

\begin{lpic}{G2_o(0.195)}
\setlength{\lpbgboxsep}{0.2mm}
\lbl[tlW]{5,165;MACSJ1115.2$+$5320 B}
\end{lpic}

\begin{lpic}{G3_o(0.195)}
\setlength{\lpbgboxsep}{0.2mm}
\lbl[tlW]{5,165;MACSJ1115.2$+$5320 C}
\end{lpic}

\begin{lpic}{H1_o(0.195)}
\setlength{\lpbgboxsep}{0.2mm}
\lbl[tlW]{5,165;MACSJ1115.8$+$0129 A}
\end{lpic}

}

\caption{Arcs detected in our sample. Each frame is a $14''
  \times 12''$ section of the HST/ACS image. The frame of MACS0451.9$+$0006 A is a $28'' \times 24''$ image section. Orientations are as in
  Fig. 1.}

\end{figure*}

\addtocounter{figure}{-1}
\begin{figure*}

\center

\makebox{
\footnotesize

\begin{lpic}{j9fc47010_B(0.195)}
\setlength{\lpbgboxsep}{0.2mm}
\lbl[tlW]{5,165;MACSJ1115.8$+$0129 B}
\end{lpic}

\begin{lpic}{I1_o(0.195)}
\setlength{\lpbgboxsep}{0.2mm}
\lbl[tlW]{5,165;MACSJ1133.2$+$5008 A}
\end{lpic}

\begin{lpic}{I2_o(0.195)}
\setlength{\lpbgboxsep}{0.2mm}
\lbl[tlW]{5,165;MACSJ1133.2$+$5008 B}
\end{lpic}

\begin{lpic}{08020_A(0.195)}
\setlength{\lpbgboxsep}{0.2mm}
\lbl[tlW]{5,165;MACSJ1149.5$+$2223 A}
\end{lpic}

}

\makebox{
\footnotesize

\begin{lpic}{J1_o(0.195)}
\setlength{\lpbgboxsep}{0.2mm}
\lbl[tlW]{5,165;MACSJ1206.2$-$0847 A}
\end{lpic}

\begin{lpic}{J2_o(0.195)}
\setlength{\lpbgboxsep}{0.2mm}
\lbl[tlW]{5,165;MACSJ1206.2$-$0847 B}
\end{lpic}

\begin{lpic}{J3_o(0.195)}
\setlength{\lpbgboxsep}{0.2mm}
\lbl[tlW]{5,165;MACSJ1206.2$-$0847 C}
\end{lpic}

\begin{lpic}{K1_o(0.195)}
\setlength{\lpbgboxsep}{0.2mm}
\lbl[tlW]{5,165;MACSJ1236.9$+$6311 A}
\end{lpic}

}

\makebox{
\footnotesize

\begin{lpic}{L1_o(0.195)}
\setlength{\lpbgboxsep}{0.2mm}
\lbl[tlW]{5,165;MACSJ1354.6$+$7715 A}
\end{lpic}

\begin{lpic}{L2_o(0.195)}
\setlength{\lpbgboxsep}{0.2mm}
\lbl[tlW]{5,165;MACSJ1354.6$+$7715 B}
\end{lpic}

\begin{lpic}{L3_o(0.195)}
\setlength{\lpbgboxsep}{0.2mm}
\lbl[tlW]{5,165;MACSJ1354.6$+$7715 C}
\end{lpic}

\begin{lpic}{L4_o(0.195)}
\setlength{\lpbgboxsep}{0.2mm}
\lbl[tlW]{5,165;MACSJ1354.6$+$7715 D}
\end{lpic}

}

\makebox{
\footnotesize

\begin{lpic}{j9en1q010_A(0.195)}
\setlength{\lpbgboxsep}{0.2mm}
\lbl[tlW]{5,165;RCS131722$-$0201.4 A}
\end{lpic}

\begin{lpic}{j9en1h010_A(0.195)}
\setlength{\lpbgboxsep}{0.2mm}
\lbl[tlW]{5,165;RCS141910$+$5326.1 A}
\end{lpic}

\begin{lpic}{j9en1h010_B(0.195)}
\setlength{\lpbgboxsep}{0.2mm}
\lbl[tlW]{5,165;RCS141910$+$5326.1 B}
\end{lpic}

\begin{lpic}{09020_A(0.195)}
\setlength{\lpbgboxsep}{0.2mm}
\lbl[tlW]{5,165;MACSJ1423.8$+$2404 A}
\end{lpic}

}

\makebox{
\footnotesize

\begin{lpic}{09020_B(0.195)}
\setlength{\lpbgboxsep}{0.2mm}
\lbl[tlW]{5,165;MACSJ1423.8$+$2404 B}
\end{lpic}

\begin{lpic}{j9en22010_A(0.195)}
\setlength{\lpbgboxsep}{0.2mm}
\lbl[tlW]{5,165;RCS151840.1$+$084500 A}
\end{lpic}

\begin{lpic}{0s010_A(0.195)}
\setlength{\lpbgboxsep}{0.2mm}
\lbl[tlW]{5,165;RCS212134$-$6335.8 A}
\end{lpic}

\begin{lpic}{10020_A(0.195)}
\setlength{\lpbgboxsep}{0.2mm}
\lbl[tlW]{5,165;MACSJ2129.4$-$0741 A}
\end{lpic}

}

\makebox{
\footnotesize

\begin{lpic}{0p010_A(0.195)}
\setlength{\lpbgboxsep}{0.2mm}
\lbl[tlW]{5,165;RCS215609.1$+$012319 A}
\end{lpic}

\begin{lpic}{11020_A(0.195)}
\setlength{\lpbgboxsep}{0.2mm}
\lbl[tlW]{5,165;MACSJ2214.9$-$1359 A}
\end{lpic}

\begin{lpic}{11020_B(0.195)}
\setlength{\lpbgboxsep}{0.2mm}
\lbl[tlW]{5,165;MACSJ2214.9$-$1359 B}
\end{lpic}
}

\caption{[Continued]}
\end{figure*}
\begin{figure*}

\makebox{
\begin{lpic}{low_z_num_hist(0.18)}
\lbl[b]{220,0;Number of arcs}
\lbl[l]{10,90,90;Fraction of clusters}
\end{lpic}
\begin{lpic}{med_z_num_hist2(0.18)}
\lbl[b]{220,0;Number of arcs}
\lbl[l]{10,90,90;Fraction of clusters}
\end{lpic}

}

\caption{Distributions of the MACS (solid line) and RCS (dashed line) clusters as a function the number
  of arcs in an individual cluster. Left panel is the low-redshift
  subsample. Right panel is the medium-redshift subsample.}
\end{figure*}
\begin{figure*}
\center
\includegraphics[width=8cm]{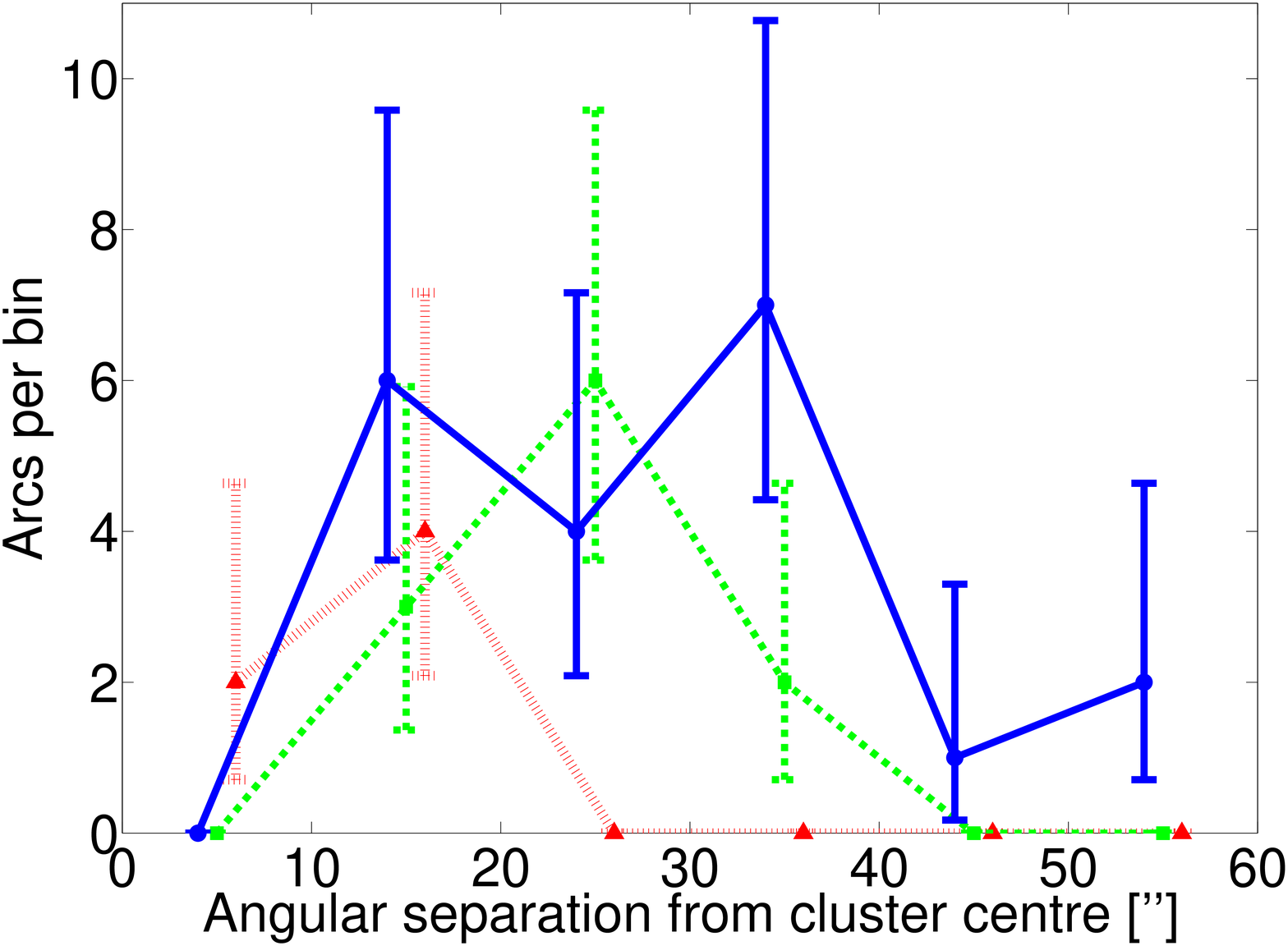}
\includegraphics[width=8cm]{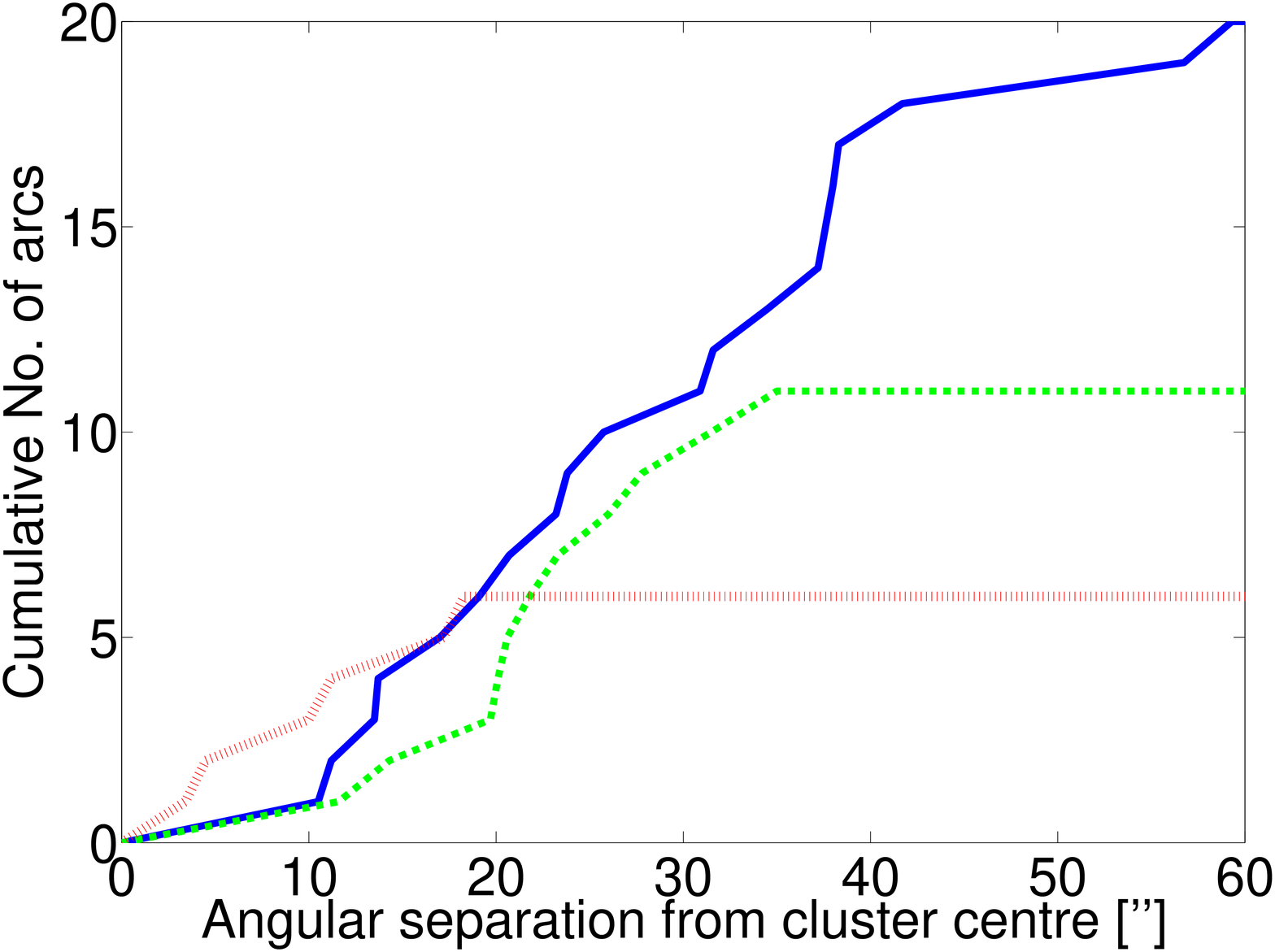}
\caption{Distribution of arc angular separations from cluster centres
  in various subsamples. Left panel is the binned distribution, right panel is cumulative. Solid (blue) line is the low-redshift MACS subsample,
  dashed (green) line is the medium-redshift MACS subsample, and dotted (red) line
  is the RCS sample (all redshifts combined). For clarity, error bars are omitted for bins
with zero arcs. We have also omitted the arc in RCS131722$-$0201.4, which is $48"$
from the centre of the cluster, as it is likely associated with a local
mass concentration near the arc rather than with the whole cluster.}
\end{figure*}

\subsection{Arc production efficiency}

Table $7$ summarizes the arc statistics of our various cluster
subsamples.
As noted above, only two arcs are detected in the RCS low-redshift
subsample, compared to the $26$ arcs detected in the low-redshift MACS
subsample. The arc production efficiencies are, therefore,
$0.11^{+0.15}_{-0.07}$, and $1.13^{+0.27}_{-0.22}$ arcs per cluster
for the RCS and MACS subsamples, respectively, where we cite a $68 \%$
confidence interval assuming Poisson statistics. In the
medium-redshift bin, the MACS clusters are also more efficient lenses
than the RCS clusters, with efficiencies of $1.33^{+0.42}_{-0.33}$, and
$0.28^{+0.19}_{-0.12}$ arcs per cluster, respectively.

With zero detected arcs, the high-redshift RCS sample has an arc
production efficiency of $< 0.24$ arcs per cluster ($95 \%$
confidence), which is consistent with the RCS efficiencies at lower
$z$. As the arc occurrence frequency is consistent among different
redshift bins, we tabulate also the total frequency in the X-ray
versus the optical subsamples. The frequencies differ at the $5\sigma$
level.

We also derive the arc production efficiencies for arcs with $l/w \geq
10$, for comparison with previous studies in which this $l/w$ ratio
was used to define giant arcs. All of the RCS arcs have $l/w \geq 10$
and therefore the RCS cluster efficiencies remain unchanged. However,
the MACS cluster production efficiencies of arcs with $l/w \geq 10$
are somewhat lowered to $0.74^{+0.23}_{-0.18}$, and
$1.08^{+0.39}_{-0.23}$ arcs per cluster for the low-, and
medium-redshift subsamples, respectively. Even so, both the low- and
medium-redshift MACS clusters are significantly more efficient lenses
($>3 \sigma$) than their RCS counterparts.
\begin{table*}
\caption{Arc statistics summary}
\smallskip
\begin{tabular}{lcccccc} 
\hline
\noalign{\smallskip}
Subsample & $N_{clusters}$ &  $N_{lenses}$ & \multicolumn{2}{c}{$N_{arcs}$} &
\multicolumn{2}{c}{Arcs per cluster}  \\
 & & & ($l/w \geq 8$) & ($l/w \geq 10$) &  ($l/w \geq 8$) & ($l/w \geq 10$)  \\
\noalign{\smallskip}
\hline
\multicolumn{6}{c}{X-ray selected clusters}\\
\hline
 XBACs ($0.17\leq z \leq 0.26$) & 10 & 7 & 17 & 12 &
 $1.7^{+0.52}_{-0.41}$ & $1.2^{+0.46}_{-0.34}$  \\
\noalign{\smallskip}
 MACS ($0.3\leq z < 0.5$) & 23 & 12 & 26 & 17 & $1.13^{+0.27}_{-0.22}$ &
 $0.74^{+0.23}_{-0.18}$  \\
\noalign{\smallskip}
 MACS ($0.5\leq z < 0.7$) & 12 & 9 & 16 & 13 & $1.33^{+0.42}_{-0.33}$ &
 $1.08^{+0.39}_{-0.23}$  \\
\noalign{\smallskip}
 MACS ($0.3\leq z < 0.7$) & 35 & 21 & 42 & 30 & $1.20^{+0.22}_{-0.18}$ &
 $0.86^{+0.19}_{-0.16}$  \\
\noalign{\smallskip}
 Total ($0.17 \leq z < 0.7$) & 45 & 28 & 59 & 42 &
 $1.31^{+0.19}_{-0.17}$ & $0.93^{+0.17}_{-0.14}$ \\
\hline
\multicolumn{6}{c}{Optically selected clusters}\\
\hline
RCS ($0.3\leq z < 0.5$) & 18 & 2 & 2 & 2 & $0.11^{+0.15}_{-0.07}$ &
$0.11^{+0.15}_{-0.07}$  \\
\noalign{\smallskip}
RCS ($0.5\leq z < 0.7$) & 18 & 4 & 5 & 5 & $0.28^{+0.19}_{-0.12}$ &
$0.28^{+0.19}_{-0.12}$  \\
\noalign{\smallskip}
RCS ($0.7\leq z \leq 1$) & 16 & 0 & 0 & 0 &  $0^{+0.12}_{-0}$ &
  $0^{+0.12}_{-0}$ \\
\noalign{\smallskip}
RCS ($0.3\leq z < 0.7$) & 36 & 6 & 7 & 7 &  $0.19^{+0.10}_{-0.07}$ &
  $0.19^{+0.10}_{-0.07}$ \\
\noalign{\smallskip}
Total ($0.3 \leq z \leq 1$) & 52 & 6 & 7 & 7 & $0.13^{+0.07}_{-0.05}$
& $0.13^{+0.07}_{-0.05}$ \\ 
\noalign{\smallskip}
\hline
\smallskip
\end{tabular} 

\end{table*}

\subsection{Notes on individual objects}

\subsubsection{Lensing signatures in clusters without giant arcs}

In addition to the automatic detection results, we have visually
inspected all of the clusters in our sample. We find that there are
several clusters that show signs of strong lensing but in which no arc
was detected algorithmically, using our detection thresholds. In some
cases, the arcs are too faint, while in others the arcs may be bright
but have a length-to-width ratio below our threshold. Sometimes, an
arc is projected close to another galaxy, making its detection
difficult. We find six such clusters with signatures of strong lensing
that are not included in our arc catalogue. Two of the
clusters (MACSJ1319.9$+$7003 and MACSJ2135.2$-$0102), are found in the
low-redshift MACS subsample, two (RCS051128$-$4235.2 and
RCS132335$+$3022.6) are in the medium-redshift RCS subsample, the fifth
cluster, MACSJ0647.7$+$7015, is in the medium-redshift MACS subsample, and
the sixth cluster, RCS RCS025242.5$-$150024, is in the high-redshift RCS
subsample. Figure 5 shows each of these cases.
\begin{figure*}
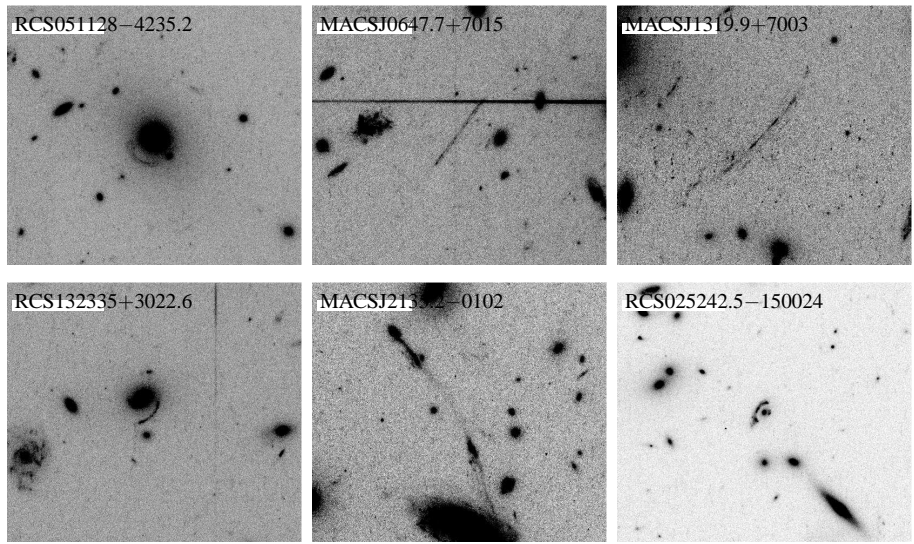


\center

\makebox{
\footnotesize

\begin{lpic}{2c010_arc(0.2)}
\setlength{\lpbgboxsep}{0.2mm}
\lbl[tlW]{5,165;RCS051128$-$4235.2}
\end{lpic}

\begin{lpic}{54020_arc(0.2)}
\setlength{\lpbgboxsep}{0.2mm}
\lbl[tlW]{5,165;MACSJ0647.7$+$7015}
\end{lpic}

\begin{lpic}{59010_arc(0.2)}
\setlength{\lpbgboxsep}{0.2mm}
\lbl[tlW]{5,165;MACSJ1319.9$+$7003}
\end{lpic}
}

\vskip 0.2cm
\makebox{
\begin{lpic}{1m010_arc(0.2)}
\setlength{\lpbgboxsep}{0.2mm}
\lbl[tlW]{5,165;RCS132335$+$3022.6}
\end{lpic}

\begin{lpic}{85010_arc(0.2)}
\setlength{\lpbgboxsep}{0.2mm}
\lbl[tlW]{5,165;MACSJ2135.2$-$0102}
\end{lpic}

\begin{lpic}{3c010_small_arc2(0.2)}
\setlength{\lpbgboxsep}{0.2mm}
\lbl[tlW]{5,165;RCS025242.5$-$150024}
\end{lpic}

}

\caption{$28'' \times 24''$ image section of arcs which were not detected
  algorithmically, using our detection thresholds, and therefore not
  included in our arc catalogue.}
\end{figure*}

\subsubsection {MACSJ1354.6$+$7715 - another bullet cluster ?}

Inspection of the image of MACSJ1354.6$+$7715 suggests the
existence of two separate galaxy concentrations. Arcs A and B (see
Fig. 1) seem to straddle one centre. About $75''$ west of that centre
there seems to be another mass concentration enclosed by arcs C and D.
We note that south of arc C there is an additional arc which is not
included in our arc catalogue due to its small $l/w$ ratio. The two
galaxies at the centres of the two concentrations have magnitudes of
19.8 mag (east clump), and 19.2 mag (west clump). This cluster may be
during some stage of a merger, but with still a considerable amount of
substructure. Although the two clumps may be chance projections of two
clusters at different redshifts, this is unlikely given the rarity of
such massive lensing clusters. Moreover, the optical colors of the
early-type galaxies across the field are also consistent with a single
redshift. Existing ROSAT data show that the X-ray emission is centred
on the system, but the entire HST/ACS field shown in Fig. 6 spans only
a few ROSAT resolution elements, making it impossible to say anything
about the X-ray flux distribution relative to the two mass and
optical-light concentrations. Higher resolution X-ray imaging (already
approved with {\it Chandra}) and
optical spectroscopy are needed to select among these alternatives.
\begin{figure*}
\begin{center}
\includegraphics[width=13cm]{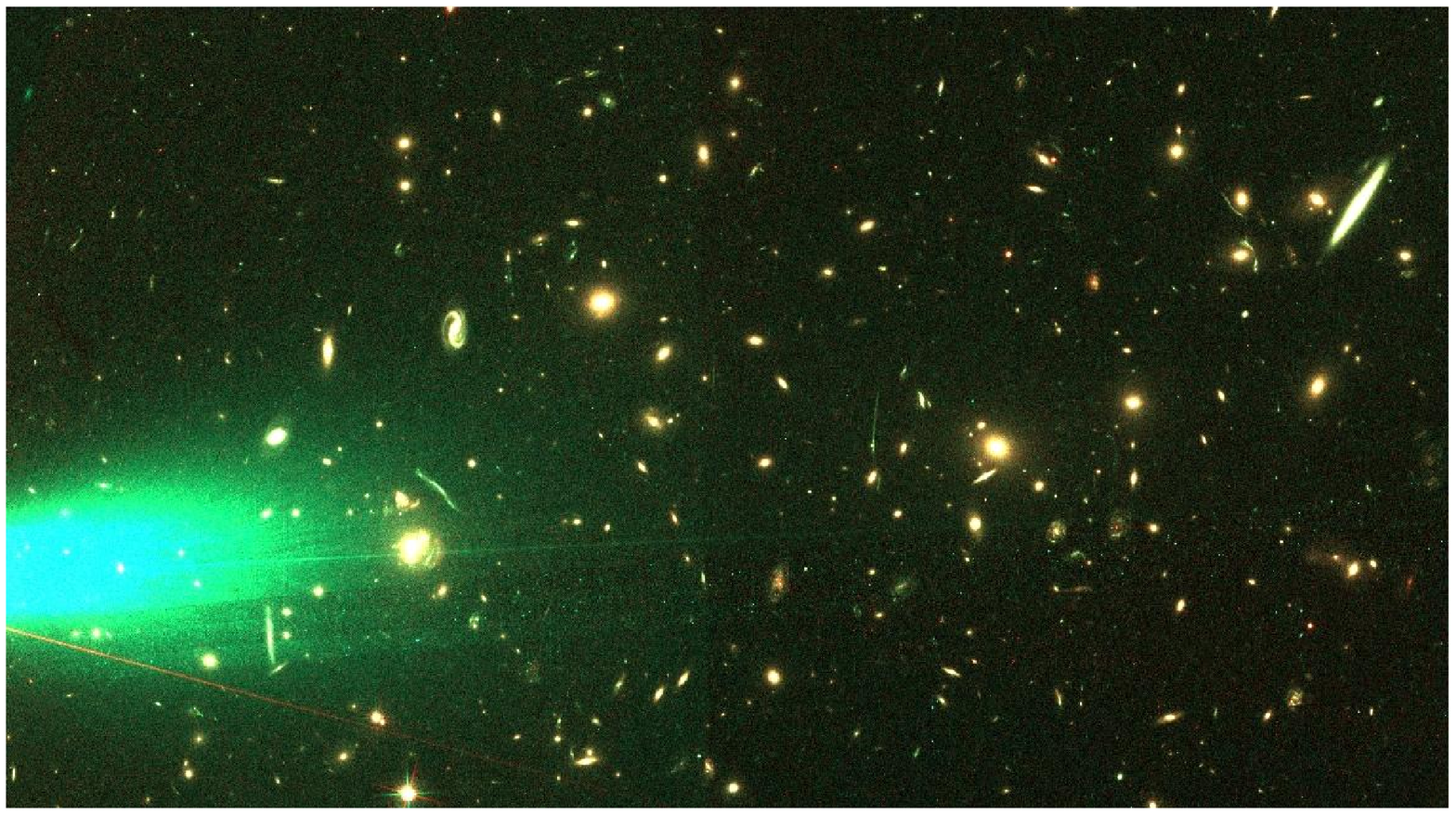}
\end{center}
\caption{A $195'' \times 105''$ image section (north is up) of the cluster
  MACSJ1354.6$+$7715. The color (in electronic version) image is a
  composite of the F606W and F814W HST/ACS images. The conspicuous feature
  on the left side is scattered light from a bright star outside
  the field of view.}
\end{figure*}

\subsubsection{Radial arcs in MACSJ2129.4$-$0741 ?}

Close inspection of the central area of the cluster MACSJ2129.4$-$0741
reveals two objects that appear to be radially distorted (See Fig.
7). The image parity of each of these objects seems to be flipped, as
expected in lensing. An alternative explanation for these objects is
tidal tails due to physical interaction between galaxies.  Again,
optical spectroscopy is needed to resolve the issue.

\subsubsection{A large arc in the field of the high
  redshift cluster RCS025242.5$-$150024.}

We have found an extraordinarily large arc ($10''$) in the cluster
RCS025242.5$-$150024 (Fig. 8). This arc is found near a galaxy which is
too bright ($m_{\rm F814W}=17.8$) to belong to this high redshift
($z=0.995$) cluster. It seems that the arc is produced by the
gravitational field of this foreground galaxy, and therefore we do not
include it in our arc catalogue.

\section{Discussion}

The results of our arc survey, presented above, can serve as a new
and improved observational basis for future arc statistic studies.
However, our survey also shows clearly that  
the arc-production efficiency of X-ray-selected clusters such
as MACS and XBACS is higher by a factor of $5-10$ than that of
RCS clusters. In this section, we carry out additional analysis and
discussion of the meaning of this result. 

At a given redshift, the cross section for lensed arc formation
depends primarily on mass, although mass profile, ellipticity and
substructure are also important. The mass dependence weakens
towards the high mass end at ${\rm M}_{\rm 200}\sim 10^{15} {\rm
  M}_\odot$ (Dalal et al. 2004; Hennawi et al. 2007). The stark
difference in the arc frequency between the X-ray selected and
optically selected clusters immediately raises the possibility that
they probe different mass ranges. Based on their X-ray luminosities,
the X-ray selected clusters have masses of ${\rm M}_{200} > 10^{15}
{\rm M}_{\odot}$. Unfortunately, there is scant information of the 
 X-ray properties of the RCS
clusters, and hence on their masses. For example,
Hicks et al. (2008) recently observed with {\it Chandra} a sample of 13 RCS
clusters, of which detailed analysis was possible for nine. They found
significant differences in the mass-temperature-luminosity relations
of X-ray selected and RCS clusters, X-ray underluminosity in some RCS
clusters,
and evidence that RCS clusters
have a larger fraction of their baryons in stars. Nevertheless, since
optical flux is one of the few observables we do have available
for the RCS clusters,
we begin by comparing the optical luminosities of the MACS and RCS subsamples. The HST/ACS field of view covers
only the central core regions of the clusters, and therefore we examine
several proxies for the optical luminosity.
   
As a first proxy for optical luminosity, we examine 
the luminosities of the brightest
cluster galaxies (BCGs) of the RCS and MACS samples.
In SDSS clusters, Hansen et al. (2005) have found a
correlation between cluster mass and BCG luminosity. 
The BCG
magnitudes were measured using SExtractor by including the light from
pixels which belong to the BCG and are above the detection threshold.
Since the low-redshift subsamples are observed through different
filters, we first calculate the F606W$-$F814W color for each RCS
cluster redshift using an elliptical galaxy spectral template from
Kinney et al.  (1996), and convert the RCS cluster BCG F814W
magnitudes to F606W. In this comparison we exclude the following
clusters (four MACS and five RCS) due to the uncertainty in
determining their centres and in identifying the dominant BCGs:
MACSJ0916.1$-$0023, MACSJ1354.6$+$7715, MACSJ2243.3$-$0935, MACSJ0257.1$-$2325, RCS131912$-$0206.9, RCS022403$-$0227.7, RCS110104$-$0351.3, RCS033414$-$2824.6, and RCS110814$-$0430.8.

We find that 
the BCG magnitudes are more uniformly distributed in the RCS subsample
than in the MACS subsample, and the BCGs span a wider magnitude range.
Nevertheless, in the low-redshift MACS and RCS subsamples, the median
BCG absolute magnitudes are, ${\rm M}_{606}=-21.9$ and ${\rm
  M}_{606}=-22$, respectively. Likewise, in the the medium-redshift
MACS and RCS subsamples the median BCG magnitudes are ${\rm
  M}_{814}=-23.8$, and ${\rm M}_{814}=-23.6$, respectively. A
Kolmogorov-Smirnov (KS) test indicates that for both the low- and the
medium-redshift subsamples, the null hypothesis that both the RCS and
MACS BCG magnitudes are derived from the same parent distribution
cannot be confidently rejected (probabilities of 0.12 and 0.35,
respectively, for the null hypothesis). These numbers are summarized
in Table 8.

For a second comparison of optical luminosities, 
we measure integrated optical luminosity of
the brightest galaxies within the cluster cores. We measure the total
light of galaxies inside a physical aperture of radius 270 kpc (at low
$z$) and 370 kpc (at medium $z$). The contribution of foreground and
background galaxies to the light is determined statistically in
annuli of $400-530$ kpc and $550-730$ kpc, for the low- and
medium-redshift subsamples, respectively, and subtracted from the core
light. The area in which we measure the ``background'' is still well
within the cluster, and hence, our cluster core luminosities are
underestimated due to background over-subtraction. Nevertheless,
barring large profile differences (see below), 
these biased estimates of cluster
luminosity can still be compared meaningfully between the X-ray and
optical samples. We include only the light from objects with
magnitudes fainter than the cluster BCG magnitude, but brighter than 24
mag. We convert the MACS low-redshift subsample's F606W luminosities to
F814W luminosities assuming, again, the Kinney et al. (1996) elliptical
galaxy template and the filter transmission curves for the two bands.
The resultant optical luminosity distributions (Fig. 9) of both the
RCS and MACS cluster are consistent with being drawn from the same
parent distribution ($0.12$ and $0.37$ probabilities for the null
hypothesis, see Table 8).
\begin{figure}
\center
\includegraphics[width=8cm, angle=0]{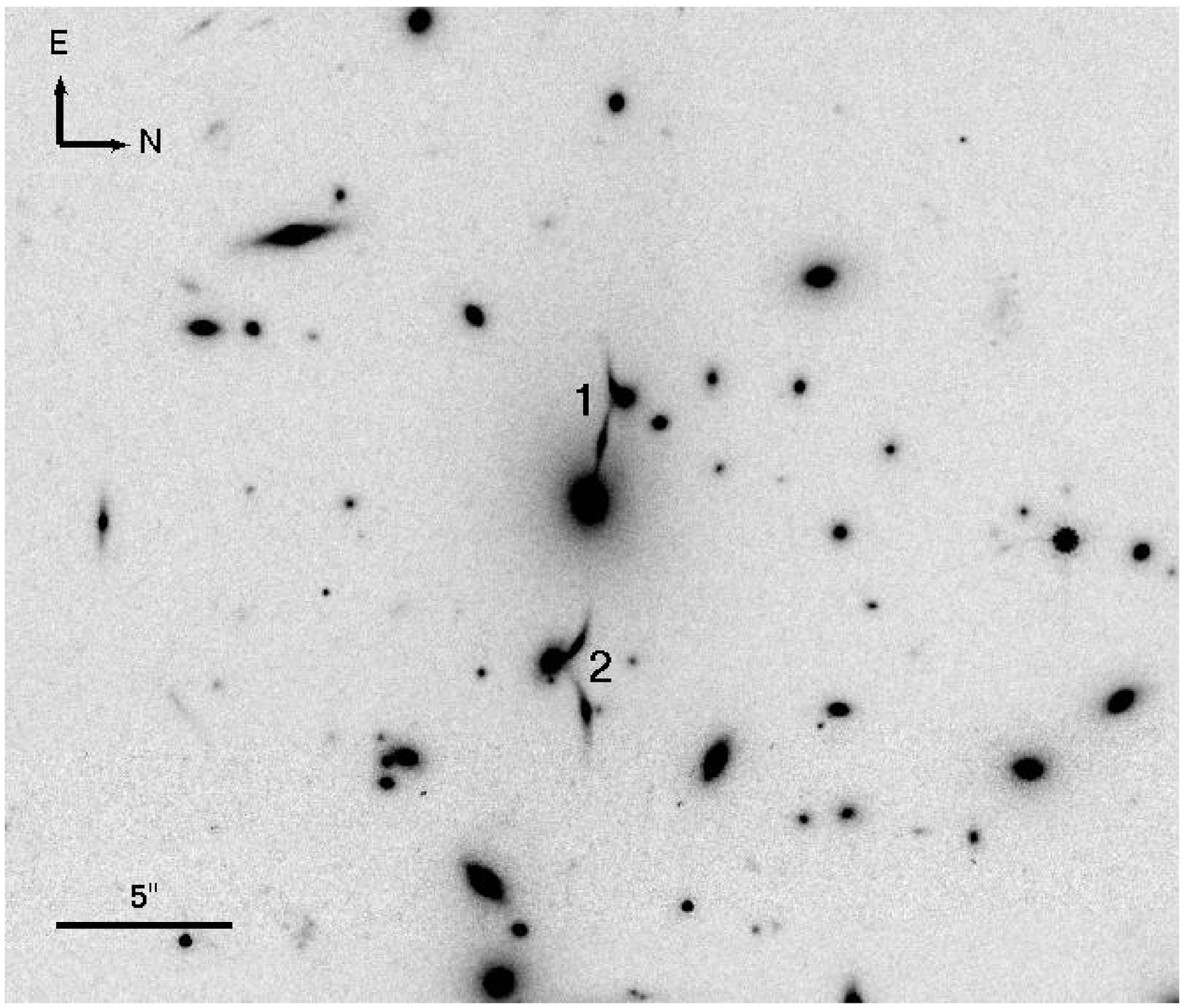}

\caption{$33'' \times 28''$ image sections of two radially distorted objects (marked as $1$ and $2$) in the
  cluster MACSJ2129.4$-$0741.}

\end{figure}
\begin{figure}
\center
\includegraphics[width=8cm, angle=0]{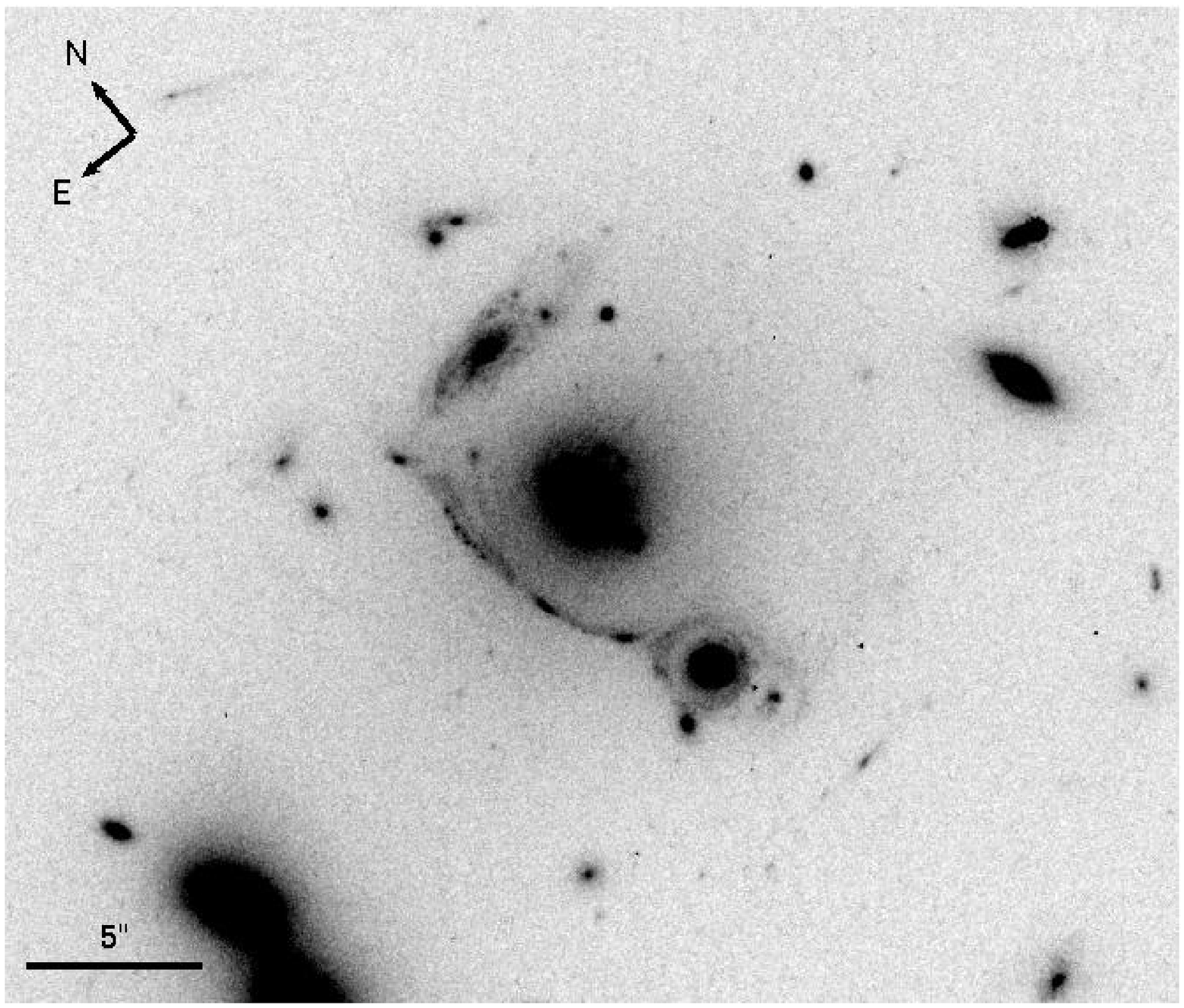}

\caption{$33'' \times 28''$ image section of a large arc found in the
  field of view of the high-redshift cluster RCS025242.5$-$150024, around
  a foreground galaxy.}

\end{figure}

Finally, as a third method of comparing optical luminosities, we
simply count the light from {\it {\bf all}} the pixels inside the
above apertures and annuli (but still leaving out the light from
objects brighter than the BCG). This method takes into account the
light from all the stars in the cluster cores, including stars in
galaxies below the detection limit and diffuse intracluster light. As
in the previous method, the core luminosity is underestimated due to
background over-subtraction, but in a consistent way for the X-ray and
optical clusters. In contrast to the two previous methods, where the
dominant galaxies in the cluster core are early types, in this case
the color correction, applied in order to convert F606W fluxes to
F814W, is less clear-cut, since fainter and undetectable dwarf galaxies
may well be blue. For the range in color terms from early-type to
late-type galaxies, the luminosity distributions are either
consistent with being drawn from the same parent distribution (0.05
probability for the null hypothesis, ``blue'' color correction), to
marginally consistent (0.01 probability for the null hypothesis,
``red'' color correction). At most (in the case of the low-redshift
subsamples, and assuming the reddest color correction) the medians of
these two distributions differ only by a factor of $1.6$.

Overall, as shown in Table 8, the medians of the RCS and MACS cluster
luminosity distributions and the results of the KS tests we applied to
these distributions suggest that the RCS and MACS cluster samples have
similar optical luminosities. We note also that the MACS clusters that
actually display arcs (shaded histograms in Fig. 9) are not
necessarily the most luminous ones, and that there is a large
overlap of their luminosities with those of RCS clusters that are much
less efficient arc producers. We have also measured and compared
 the optical light profiles of the two samples and, within the limited
 range of the cluster cores covered by the ACS data, we find no
 significant differences. 

\begin{table*}
\caption{Comparison of MACS and RCS cluster luminosities}
\smallskip
\begin{tabular}{lcc} 
\hline
\noalign{\smallskip}
Subsample & Optical luminosity measure &
KS Probability  \\
\noalign{\smallskip}
\hline
\multicolumn{3}{c}{Cluster BCG absolute magnitudes}\\
\hline
 MACS ($0.3\leq z < 0.5$) & $-21.9$ & \multirow{2}{*}{$0.12$} \\
 RCS  ($0.3\leq z < 0.5$) & $-22$ & \\
 MACS ($0.5\leq z < 0.7$) & $-23.8$ & \multirow{2}{*}{$0.35$} \\
 RCS  ($0.5\leq z < 0.7$) & $-23.6$ &\\ 
\hline
\multicolumn{3}{c}{Cluster core luminosities (light of bright galaxies
only)}\\
\hline

 MACS ($0.3\leq z < 0.5$) & $1.5$ & \multirow{2}{*}{$0.12$} \\
 RCS  ($0.3\leq z < 0.5$) & $1.3$ & \\
 MACS ($0.5\leq z < 0.7$) & $2.4$ & \multirow{2}{*}{$0.37$} \\
 RCS  ($0.5\leq z < 0.7$) & $2.2$ & \\
\noalign{\smallskip}
\hline
\multicolumn{3}{c}{Cluster core luminosities (total light)}\\
\hline
 MACS ($0.3\leq z < 0.5$) & $3.8$ & \multirow{2}{*}{$0.01$} \\
 RCS  ($0.3\leq z < 0.5$) & $2.3$ & \\
 MACS ($0.5\leq z < 0.7$) & $4.5$ & \multirow{2}{*}{$0.07$} \\
 RCS  ($0.5\leq z < 0.7$) & $3.1$ & \\

\noalign{\smallskip}
\hline
\smallskip
\end{tabular} 

Note- Optical luminosity measure indicates F606W or F814W absolute magnitudes for the BCGs, and
luminosities in units of $10^{45}~{\rm erg}~{\rm s}^{-1}$ for the two measures of
core optical luminosity. Probability is for the null hypothesis that a
pair of distributions are not different.

\end{table*}

However, stars, let alone the small fraction
 of the stars that dominate the optical luminosity, are a tiny
 component of the total cluster mass, and it is therefore
 plausible that the masses of the two samples are very different, despite
the similar optical luminosities, 
with masses significantly below $10^{15} {\rm M}_\odot$ for the RCS clusters.
A strong argument for such a mass difference
 is the difference in the space densities
of the two samples. From the numbers of clusters and the area surveyed
(see \S~2), the projected density of MACS clusters is
$\sim 0.01 {\rm ~deg}^{2}$. Assuming the cluster mass function is
 probed correctly by X-ray surveys,  
only about one MACS-like massive cluster  is expected in
the $\sim 100 {\rm ~deg}^{2}$ search area of the RCS survey.
Based on the cluster mass function (e.g. Reiprich \& B{\" o}hringer 2002), 
the $\sim 1000$ clusters 
found in the RCS search area imply that the vast majority of these
clusters have masses   of ${\rm
  M}_{\rm 200} = 10^{14} {\rm M}_\odot$, an order of magnitude lower than MACS
 clusters. 

This picture is further supported by the distributions of arc
separations from the cluster centres. Since arcs occur near critical
curves, the separations can roughly represent the Einstein
radii of the clusters. As noted in \S3 and seen in Fig.~4, the MACS
clusters have arcs at $10''-50''$, with a median at $24''$, while
the RCS arcs are generally much closer in, with a median of $10''$.
The small Einstein radii of most of the RCS clusters with arcs 
are similar to those of rich groups.

A puzzling corollary of the above arguments, however, is the fact that 
in the small subsample of 52 RCS clusters imaged with ACS, which 
constitute  just 5 per cent of the full RCS sample, there are as many as 
two clusters
(RCS 141910$+$5326.1 and RCS 215609.1$+$012319) with arcs at separations
implying $20''$ Einstein radii, and hence MACS-like masses, in
contrast to the expectation that of order just one such cluster exists 
in the {\it entire}
RCS survey. Furthermore, despite their mass, only about half of the
MACS clusters display arcs in our survey, because a galaxy in a
suitable position in the source plane is required in order to produce
an arc. The two
large-separation RCS clusters in the HST sample would thus imply about
4 massive RCS clusters in the HST sample, and $\sim 100$ in the full
RCS sample, as opposed to the $\sim 1$ expected from the X-ray-derived
mass function. 
A possible explanation is that the HST RCS sample is
not a fully representative selection of the RCS. Indeed, the wide arcs 
of the cluster RCS 141910$+$5326.1 above were already noted in ground-based
images by Gladders et al. (2003), and it may have been included in
the HST sample for this reason. Thus, the HST RCS sample could be a
representative subsample of the RCS, {\it plus} a few of the most
massive RCS clusters, and would thus be pre-biased in favor of
lensing. Since, despite this bias, the X-ray-selected clusters are
still much more efficient lenses, the observed arc occurrence frequency in
the RCS clusters imaged with HST provides an upper limit on
the RCS arc frequency as a whole.
   
On the other hand, Gladders et al. (2003) discussed eight potential lenses
out of the full RCS sample of about 1000 clusters. With random selection,
one would expect 1.1 of these eight lenses to be included among the 150 RCS
clusters in the HST Snapshot sample, out of which the actually
observed targets were chosen by HST schedulers. In fact, there are two
of the eight Gladders et al. (2003) potential lenses among the 150 HST
targets. This could be the result of some slight bias in favor of
known lenses, as described above, but could of course be due just to
chance. We reiterate that, if the RCS sample is unbiased, our conclusion
about the relatively low lensing inefficiency of RCS clusters hold. 
If the RCS sample was pre-biased, this conclusion is only strengthened.

 The simplest explanation for our measurement of a low lensing
 efficiency among RCS clusters, compared to X-ray-selected clusters, is 
 a typical RCS cluster mass that is lower by an order of
 magnitude. This leaves open the question of what stands behind the
 similarity of X-ray and RCS clusters, in terms of stellar luminosity,
 optical profiles, numbers of galaxies, and general optical appearance. These
 similarities cannot be due to chance line-of-sight projections 
 (RCS clusters are
 chosen based on redshifts of the early-type galaxies that characterise
 dense environments, so they are, in fact, real associations),
 nor due to projection effects along large-scale structure
 ``filaments'' -- simulations have shown that large scale structure may contribute only about 10 per cent
 to the cluster surface mass density (Wambsganss, Bode, 
 \& Ostriker 2005; Hilbert et 
 al. 2007). Instead, there is a real and large variation in the 
 total-mass-to-optical-light ratio among clusters.
The low mass-to-light ratio of RCS cluster cores may be caused by a
bias in favour of line-of-sight mergers in the optical selection
 process, 
a prominent and spectacular example of which is Cl0024+24 (Czoske et
 al. 2002). 
Indeed, extensive spectroscopic follow-up of RCS clusters has
 uncovered several cases 
of close projection effects of possibly physically associated systems
 as well as 
line-of-sight substructure (Gilbank et al. 2007; Cain et al. 2008). 

 A further effect to consider is the question of
 whether X-ray selection may favour the inclusion of clusters that are
 in the process of merging. Torri et al. (2004) have found that, during
 a merger, the lensing cross section is increased by a factor of $5-10$
 for a duration of a couple of hundred million years, while the X-ray
 luminosities of merging clusters are increased by a factors of $\sim
 5$. A similar conclusion regarding the X-ray luminosity of clusters
 during mergers was reached by Randall, Sarazin, \& Ricker (2002). If
 X-ray-selected cluster samples indeed have a larger fraction of
 merging clusters, one could thus expect a larger fraction of highly
 efficient lenses in those samples. Thus, the masses of the X-ray-selected 
 clusters may be systematically overestimated as well.

An interesting question is whether comparable optically and X-ray-selected cluster samples at $z > 0.7$ also differ in their arc
production efficiencies. We did not find giant arcs in any of the
high-redshift RCS clusters we analyzed, even
though their optical luminosities are comparable to those of the RCS
clusters at low and medium redshifts. This contrasts with the results
of Gladders et al. (2003) who found RCS clusters to be more efficient
lenses at high redshift.

Finally, the many arcs found in the MACS low- and medium-redshift subsamples
provide a statistically improved handle on the angular
distribution of arcs in clusters. Our results show that arcs do form
at large angular separations from cluster centres, at up to $60''$, in
some cases. Thus, the large Einstein radius of Abell 1689 is probably
not unique.

\section{Summary}

We have conducted an algorithmically based search for lensed arcs in
$\sim 100$ clusters observed with HST. Our cluster sample includes an
X-ray selected subsample (XBACs; MACS) and an optically selected
subsample (RCS), each in a range of redshifts. Our search for giant
arcs has produced $12$, $17$, and $13$ arcs ($l/w > 10$) in the XBACs,
MACS low-redshift, and MACS medium-redshift subsamples, respectively.
Only $2$, $5$, and zero arcs were found in the low-, medium-, and
high-redshift RCS subsamples. The arc production efficiency of the
MACS clusters is therefore higher by a factor of $5-10$ than that of
the RCS clusters. The typical Einstein radii of MACS clusters are
several times larger than those of the relatively few RCS cluster
that do display strong lensing. If, as we suspect, the HST sample of
RCS clusters was pre-selected in a way that favored strong lenses,
then these conclusions would only be strengthened.  

These results constitute direct evidence, based on strong-lensing
statistics, that optically selected RCS clusters are an order of magnitude
less massive than X-ray selected clusters, despite their similar
optical properties. This conclusion is supported by the factor-100
higher space density of RCS clusters. In the arc statistics literature to date, the observed statistics from
X-ray and optical clusters have often been discussed
together and interchangeably. We have demonstrated that
X-ray and optically selected clusters likely probe distinct parts
of the cluster mass function, and
should therefore
not be mixed in this way.

In a forthcoming paper, we will address arc statistics from a
theoretical point of view. We will present strong lensing statistics 
predictions using
clusters from several of the latest cosmological simulations,
calculated specifically for comparison with the observed samples
analyzed here. We expect that the observational database we have
presented,
 compared to these improved new simulations, will elucidate some
of the contradictions that have been encountered to date in this field.
\begin{figure*}
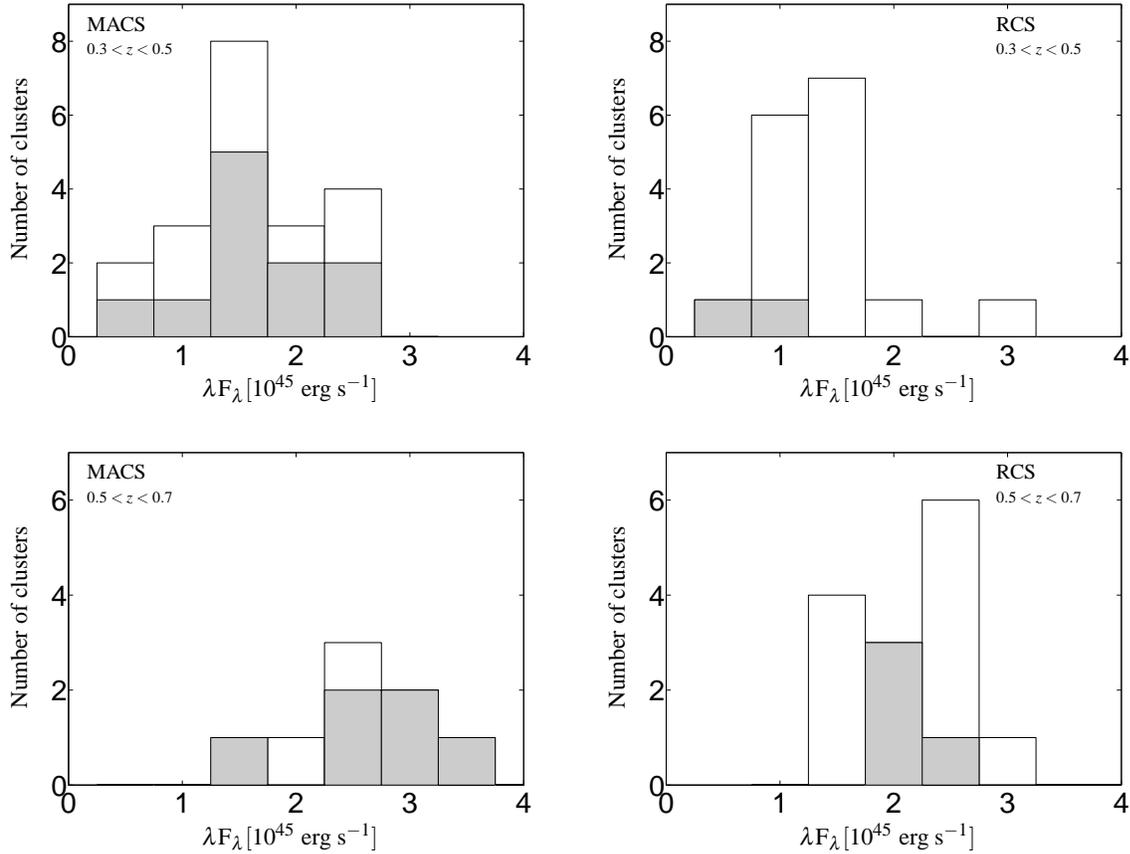

\makebox{
\begin{lpic}{macs_low_obj_hist2(0.18)}
\lbl[tl]{70,270;{\footnotesize MACS}}
\lbl[tl]{70,250;{\scriptsize $0.3 < z < 0.5$}}
\normalsize
\lbl[b]{220,-15;$\lambda{\rm F}_{\lambda}[10^{45}~{\rm erg}~{\rm s}^{-1}]$}
\lbl[l]{20,90,90;Number of clusters}
\end{lpic}
\begin{lpic}{rcs_low_obj_hist2(0.18)}
\lbl[tl]{300,270;{\footnotesize RCS}}
\lbl[tl]{300,250;{\scriptsize $0.3 < z < 0.5$}}
\normalsize
\lbl[b]{220,-15;$\lambda{\rm F}_{\lambda}[10^{45}~{\rm erg}~{\rm s}^{-1}]$}
\lbl[l]{20,90,90;Number of clusters}
\end{lpic}
}
\vskip 0.5cm
\makebox{
\begin{lpic}{macs_med_obj_hist2(0.18)}
\lbl[tl]{70,270;{\footnotesize MACS}}
\lbl[tl]{70,250;{\scriptsize $0.5 < z < 0.7$}}
\normalsize
\lbl[b]{220,-15;$\lambda{\rm F}_{\lambda}[10^{45}~{\rm erg}~{\rm s}^{-1}]$}
\lbl[l]{20,90,90;Number of clusters}
\end{lpic}
\begin{lpic}{rcs_med_obj_hist2(0.18)}
\lbl[tl]{300,270;{\footnotesize RCS}}
\lbl[tl]{300,250;{\scriptsize $0.5 < z < 0.7$}}
\normalsize
\lbl[b]{220,-15;$\lambda{\rm F}_{\lambda}[10^{45}~{\rm erg}~{\rm s}^{-1}]$}
\lbl[l]{20,90,90;Number of clusters}
\end{lpic}
}
\caption{Distributions of the MACS (left) and RCS (right)
  core optical luminosities at low redshift (top) and medium redshift
  (bottom) measured using galaxies brighter than $m_{\rm F814W}=24$.
  Shaded histograms designate the clusters that display one or more
  lensed arcs.}
\end{figure*}
 
\section*{Acknowledgements}
We thank Eran Ofek and Dovi Poznanski for their help with synthetic photometry and Keren Sahron for useful discussions. We thank the anonymous referee for constructive comments. D.M. acknowledges support by the Israel Science Foundation and by the DFG through German-Israeli Project Cooperation grant STE1869/1-1.GE625/15-1.
This research has made use of NASA's Astrophysics Data System (ADS) Bibliographic Services, as well as the
NASA/IPAC Extragalactic Database (NED). This work is based on observations made with the NASA/ESA Hubble Space Telescope, obtained from the data archive at the Space Telescope Science Institute. STScI is operated by the Association of Universities for Research in Astronomy, Inc. under NASA contract NAS 5-26555. HE gratefully acknowledges financial support from STScI grants GO-09722, GO-10491, and GO-10875. MB was supported by the Transregio-Sonderforschungsbereich TR 33 of the 
Deutsche Forschungsgemeinschaft.

{}

\end{document}